\documentclass [12pt]{article}
\usepackage {amssymb}
\usepackage {amsmath}
\usepackage{graphicx}
\usepackage{cite}
\usepackage {longtable}

\title{Physics of Limit Values at Planck scale}

\author{$^{1}$\textbf{Yu.L.Bolotin}, $^2$\textbf{A.V.Tur}, $^{3,4}$\textbf{V.V.Yanovsky}}

\begin{document}

 \maketitle

$^{1}$National Science Center "Kharkov Institute of Physics and Technology",
1, Akademicheskaya str., 61108, Kharkov, Ukraine

$^{2}$\textit{Universit\'{e} de Toulouse [UPS], CNRS, Institut de Recherche en Astrophysique et Plan\'{e}tologie,
9 avenue du Colonel Roche, BP 44346, 31028 Toulouse Cedex 4, France}

$^{3}$ \textit{Institute for single crystals, National Academy of
Science of Ukraine, Nauki Ave 60, 61001 Kharkiv, Ukraine}

$^{4}$\textit{V. N. Karazin Kharkiv National University, 4 Svobody Sq., Kharkiv, 61022, Ukraine}

\abstract{The traditional formulation of the ultimate goal of physics (in the narrower sense of axiomatic theory) involves the derivation of physical laws from first principles. Though, such option doesn’t make things easier since the task of the first principles finding is not less complicated versus to the original problem. The alternative path for understanding the world around us is to interpret the fundamental limit values as a factor determining the physical laws structure. A significant part of this path has already been completed. It was possible to show that the quantum mechanics can be built on the basis of the existence of the minimum quantum action, while the special theory of relativity - on the maximum speed $c$. Furthermore, from rather recently it became clear that a similar approach could be implemented in general relativity but in this case it can be constructed by postulating the existence of a minimum lengths. The goal of this review is to demonstrate the effectiveness of limit values as a tool for describing the physics of the Planck scale. Moreover, by virtue of their universality, the limit values allow us to establish relationships between, on first glance, distant fields of physics. We will consider the simplest consequences of the inclusion of gravitational effects in quantum reality. The most important consequence of this consideration is the inevitability of transition from the classical concept of continuum to the concept of the discrete space-time.The new physics generated by such transition will be in the center of our attention.}

\tableofcontents

\section{Introduction}

The focus of our attention will be physics of the Planck scale. Let us characterize more in detail the object of this study. Three fundamental world constants which are the gravitational constant $G$, the speed of light $c$ and Planck constant $\hbar$  allow us to construct so called Planck units of mass, length and time
\[{m_{Pl}} = \sqrt {\frac{{\hbar c}}{G}}  \simeq 2.18 \times {10^{ - 8}} \, kg,\]
\[{l_{Pl}} = \sqrt {\frac{{\hbar G}}{{{c^3}}}}  \simeq 1.6 \times {10^{ - 35}}   \, m,\]
\begin{equation}\label{e1_1}
t_{Pl} = \sqrt {\frac{\hbar G }{c^5}}  \simeq 5.39 \times 10^{- 44} \, sec
\end{equation}
In his report to the Prussian Academy of Sciences in Berlin (1899), introducing new universal constants, Planck noted that these units retain their natural significance, as long as the laws of gravitation and thermodynamics are valid,(besides  $\hbar$, $c$ and $G$ Planck included in the number of universal constants the Boltzmann constant also) \cite{1s}.

The introduction of the Planck system of units even before the creation of quantum mechanics, special and general relativity has become a significant event in physics. System "natural units" (as M.Planck called the system) raised questions far ahead of the current state of physics. To solve these problems, significant progress in physics was required. After more than a century, we are only groping for approaches to the problems posed. Final understanding will come perhaps only after the creation of the quantum theory of gravity. Systems of units based on various fundamental constants, not only reflect the history of the development of physics, but they allow us to judge the prospects for its development. In this sense, the value of Planck units still remains one of the potential clues to look into the future.

However, at first glance, it seems that Planck’s time and length are so small, and Planck's energy ($E_{Pl}=1.2 \times 10^{19} GeV$)  is so great that even in the distant future it is extremely unlikely to detect physics of such magnitude. "There is no place for such numbers in physics. They are ridiculously unnatural" - stated P. Bridgman (Nobel Laureate of 1946). But now let us return, to the Planck scale. Just very recently, inconceivable recklessness was required to even mention the possibility of experimental studying Planck scale physics (PSP). Nevertheless, contrary to the widespread "public opinion" PSP shifts in the center of attention of both theorists and observers \cite{2s,3s,4s,5s,6s,7s,8s,9s,10s,11s,12s}. The reason for this concernment is that a Planck scale gravity (the weakest of the four known interactions) turns out to be competitive. This circumstance immediately raises the problem of the synthesis of quantum mechanics with the general relativity. It is generally accepted that the ultimate goal of such synthesis is creation of the theory called quantum gravity. However, even the need to build such theory is doubtful.

Denial arguments, in the most crucial form formulated by F. Dyson \cite{13s} are: "Quantum gravity is physically meaningless. Any theory of quantum gravity implies a particle "graviton" (quant of gravity), just like a photon (quant of light). The presence of photons is easy to detect by electrons knocked from a metal surface under the action of light. But the gravitational interaction is incredibly weaker than the electromagnetic one. Therefore, the characteristic times of such effects exceed the age of the Universe. Hence, if experimentally individual gravitons cannot be detected, they have no physical reality". Let's give this statement quantitative meaning.

The lifetime of a hydrogen atom in the 3d state with respect to decay into the $1s$ state due to gravitational radiation is $\tau  \sim \times {10^{38}} \, \sec $  while Universe age ${T_{Universe}} \sim {10^{17}} \, \sec$. For optical transitions, the average lifetime  $\tau  \sim {10^{ - 7}} - {10^{ - 8}} \, \sec $.

In a more general formulation, the difficulty of combining quantum mechanics and general relativity is related to the discreteness of the first and the continuity of space-time of the second. Physicists not for the first time have uncounted such a problem, so it’s useful to turn to historical experience \cite{14s}.

As it is known, the energy spectrum of any quantum system performing a finite motion is always discrete. Indeed, according to the uncertainty principle, a single quantum state cannot occupy a phase volume $V_1 \leq \hbar^N$  where $N$  is the dimension of the configuration space. Thus, the movement bounded by the volume $V$  will contain $V/V_1$  eigenstates. According to ergodic theory, such a motion is considered as regular, in contrast to chaotic motion with a continuous spectrum and exponential instability of phase trajectories.

This result can be briefly stated in the spirit of the principle of complementarity: classical evolution is determined but random, quantum evolution is nondeterministic and nonrandom. In other words, the problem is that the discrete nature of the spectrum will never allow chaos in any quantum system with finite motion. In other words, the problem is that the discrete nature of the spectrum negates the appearance of chaos in any quantum system with finite motion. Meanwhile, the correspondence principle requires chaos in the semiclassical limit. If we agree with the point of view that chaos is impossible in quantum mechanics, (then) it would be logical just to refuse from further study of this issue. But that will mean that we shy away from the challenge Nature has given us when the limit of small   and large times exists. This is equivalent to ignoring such fundamental phenomena as turbulence or phase transitions.

An alternative point of view is not wait to the final solution for this problem but study limited version. That is study the special properties of quantum systems whose classical analogues are chaotic \cite{15s}.

A similar tactic has recently been applied to the issue of unification. quantum mechanics and general relativity. Instead of trying to construct a theory of quantum gravity, considerable efforts have been made to search for quantum manifestations of classical gravity. This area is often called the phenomenology of the Planck scale physics.

Contrary to popular belief, the problem is not that gravity cannot be quantized, but in the fact, that there are too many quantization methods and not one of them is satisfactory. Expanding the base of phenomenology, we can facilitate the choice of an adequate quantum theory.

The main difficulty of PSP is extreme parameters of studied objects. These parameters are many orders of magnitude higher than the achieved measurement limits. Thus, for protons accelerated by LHC $E \sim {10^4}\, GeV$ and therefore
\begin{equation}\label{e1_2}
  \frac{E}{{{E_{Pl}}}} = \frac{{{{10}^4}}}{{{{10}^{19}}}} = {10^{ - 15}}
\end{equation}
The linear accelerator at which Planck energies can be reached must exceed the size of the galaxy. Current accuracy of measurements of lengths are just as far from the Planck scale. Interferometers designed to detect gravity waves, have a record accuracy $\Delta \sim {10^{ - 16}} \, cm$, hence
\begin{equation}\label{e1_3}
  \frac{\Delta }{{{l_{Pl}}}} = \frac{{{{10}^{ - 16}}}}{{{{10}^{ - 33}}}} = {10^{17}}
\end{equation}
However, these estimates should not cause pessimism. The first steps are being taken in number of areas of experimental researches.
\begin{enumerate}
  \item The Early Universe \cite{16s,16ss}
  \item Disasters of cosmological scales \cite{17s}
  \item Quantum optics \cite{18s}
  \item Analog gravity \cite{19s}
\end{enumerate}
Note that analog gravity is a research program that studies the analogs of GR in other physical systems (as a rule, but not exclusively in condense matter) in order to obtain a new understanding of quantum manifestations of classical gravity.

It is easy to see that the functional roles of the constants $\hbar$, $c$ and $G$  used to construct Planck units are is different. While If the first two represent the limit values (minimum quantum of action and maximum speed of interaction propagation) and underlie quantum mechanics ($hbar$) and special relativity ($c$), then the Newtonian constant $G$ "only" fixes the magnitude of the gravitational interaction. In other words, the Newtonian constant $G$ does not fulfill any restrictive functions. Therefore, it seems natural to make the set of fundamental constants more uniform and more efficient for subsequent analysis of physics on the Planck scales, replacing the gravitational constant $G$ by some new limit value associated with GR. In this case, Planck units will be expressed only through fundamental limit values that underlie quantum mechanics ($\hbar$), SR ($c$) and GR (this quantity we must determine).

Limit values play a central role in both phenomenology and axiomatic (microscopic) theories. Assertion of the existence of limit values can be used as the basis of physical axiomatics. Well known that quantum mechanics can be built by postulating the existence of minimal quant of action, SR - SRT - by limiting the speed of propagation interaction. Relatively recently, it became clear that a similar approach which can be implemented to GR: theory can be constructed by postulating the existence of a maximal power (force) \cite{20s,21s,22s}.

The limit values apply to any physical systems, regardless of their nature and for any observer. The concrete magnitude of the limit value is not so important: the decisive role is played by the fact of its existence. Let us clarify that the concept of limit value differs from the concept of record: record can be improved; whereas it is impossible for limit value.

By fundamental limit values we mean those ones which cannot be deduced from existing theories, and existence of which can be basis of future axiomatics theories. Status of fundamental limit value can be achieved in the process of physical evolution. The classic example of this is the speed of light: formal parameter (Maxwell), electromagnetic wave speed (Hertz), limit speed of interaction transfer (Einstein).

The goal of this review is, firstly, to demonstrate the effectiveness of limit values as a tool for describing the physics of the Planck scale. Secondly, to illustrate that by virtue of their universality, the limit values allow us to establish relationships between, on first glance, distant fields of physics.

\section{Discreteness and continuity}

Choice between these ideas, apparently, will have to be implemented when combining the two great theories GR and quantum theory in quantum gravity. In order to implement this, you need either, translate GR to a discrete level or a quantum theory to a continuous one. We will return to discussion of this problem below.

Let's start with discreteness. This idea arose from divisibility into parts say a bar of gold. Following Democritus, and he followed even more ancient sources, take a bar of gold and divide it in half. We get two bars, but smaller. Now take one of them and divide it in half again. Repeat this procedure many times until we get to such a bar, after dividing which we do net get two smaller bars of gold, but it will be hell knows what. This will indicate the existence of a minimum size object. Democritus called it the atom. It's clear that this reasoning can be applied to any object. Using this gedanken procedure one can establish the minimum sizes of arbitrary objects.

But what happens if this procedure is not interrupted on any scale? Then we will be forced to recognize the corresponding essence continuous or, in other words, not having discrete nature. Such a gedanken procedure of infinite division, let it be a segment of a path, is fraught with the demon of infinity in itself.

With the concept of infinity, a number of paradoxes are associated. Zeno (V-th century BC) was the first who drew attention to this issue He formulated several paradoxes (or aporias)\footnote{from Greek $\alpha \pi o \rho \iota \alpha$ – query (puzzle)  Only 9 out of his 40 aporias have survived to our time.}   which were destined to survive more than two millennium and have not lost their relevance even now. An eminent mathematician Bertrand Russell wrote: "These paradoxes in one form or another affect the foundations of almost all theories of space-time, offered from his time to the present day". Moreover, they even nip on ahead the paradoxes of modern set theory.

Generally accepted that the starting point of the crisis of the foundations of mathematics was Bertrand Russell’s letter to a mathematician by name Frege, in fact, this crises is rooted in the ideas of Aristotle himself.

That’s what he wrote in his book "Metaphysics": "Infinity is always in possibility, not in reality". It was this phrase that Russell quoted in his letter. This quote concerned such a fundamental concept as Infinity! Since the days of Aristotle, two approaches to the concept of infinity have been known. As seen from the quote above, only potential infinity was admitted by Aristotle. The idea of potential infinity has the ability always to increase the available final list of objects. Actual infinity covers is all an infinite set of objects. It contains all the objects of this set. Thus, there is nothing to add to this set.

As early as 1638, Galileo made his contribution to the justification of the impossibility of actual infinity in his book "Conversations and mathematical proofs regarding two new sciences"\footnote{It is interesting to note that exactly Statics and Dynamics were meant under those two new sciences in the Galileo’s book.}.  Amazingly, that his reasoning anticipated some of Cantor's ideas. Galileo compared two infinite sequences of numbers. The first one  is a natural series of numbers $1,2,3, \ldots$, and the second is an endless series of squares $1,4,9, \ldots$
\begin{itemize}
  \item $1 \leftrightarrow 1$
  \item $2 \leftrightarrow 4$
  \item $3 \leftrightarrow 9$
  \item $4 \leftrightarrow 16$
  \item $5 \leftrightarrow 25$
  \item $\ldots \leftrightarrow \ldots$
\end{itemize}
By setting a one-to-one correspondence between the sequence members, Galileo concludes that the number of numbers is the same in these sequences, what was a contradiction for him. In Proving this, he relied on the principle of Aristotle, that "the whole is always greater than any part of it". But in his case, the part coincides with the whole, which means a contradiction. Thus, Galileo comes to the absurdity of the actual infinity.

Fascinatingly, that 250 years later, G. Cantor, repeating these arguments, came to the opposite  conclusion, by rejecting the principle of Aristotle - the whole is more than a part. It is Cantor who introduced into mathematics actual infinity. And so it was, that, the theory of infinite sets was created.

Let us return now to the aporias of Zeno. The most famous one is Achilles and the tortoise. In this aporia, Zeno points out the difficulty in understanding of the movement. This aporia is formulated as follows: Achilles will never catch up with turtle, because they will always be separated by a finite distance. In other words, it is impossible to carry out an infinite number of actions in a finite time. The inexhaustible cannot be exhausted.

Zeno understood that a contradiction arises even in the absence of motion. To demonstrate this, he proposed an aporia of a dichotomy. According to this aporia, Achilles will not even be able to start moving. Let him need to go through some finite path. In order to pass it, he must first go half the way, but in order to go half, he must go half a half, and so on ad infinitum.

The source of these contradiction is the infinite divisibility of space and time. It is hard to imagine the feasibility of an infinite number of actions in a finite time. Of course, the problem is not in the convergence of the series, as is often believed, but in the absence of the last step. Over more than 2 millennia, a huge number of discussions, explanations, refutations and interpretations of aporias have been accumulated and we will not deal with their analysis. What matters for us  that the infinite divisibility of space and time is not trivial and can serve as a source of problems. There is no doubt that mathematical analysis easily copes with the mathematical problems of motion in abstract continuous space. The main difficulty is matching the mathematical model and real space-time. How real motion occurs and what is the fundamental structure of space-time still remains a deep problem.

Strictly speaking, dividing objects into many disjoint subsets is fraught in itself, if not paradoxes, then amazing difficulties for our intuition. The brightest demonstration of this is the famous Banach- Tarski theorem \cite{zen1}. Without going into the exact formulation that can be gleaned from \cite{zen2}), for clarity, we restrict simple illustration. According to this theorem, a ball the size of our planet can divided into a finite set of pairwise disjoint parts, of which, by translation and turns you can collect a ball the size of a billiard ball and put it in your pocket. At first, this result was perceived as a Banach-Tarski paradox. Suspicion fell on the axiom of choice. However later her innocence was proven. Now this paradox is resolved and the reason is the existence of immeasurable sets. Basically this the result should not be very surprising, if you agree that a unit segment contains as many points as a segment a million times longer. Needless to say, to carry out this procedure with a real billiard or better golden ball physically impossible. The required division into parts cannot be realized due to discrete structure of matter.

Infinity creates problems even in classical mechanics \cite{zen3}. To demonstrate them we can consider the collision of a point particle of mass $m$  with an infinite sequence of the same particles located on a unit segment $[0, 1]$. Let be the position of the  $i$-th particle $x_i =1/2^i$  where $i = 0, 1, 2, 3,\ldots$. Now the particle with $i=0$  which is located at the point $x=1$ we give the velocity $v$ towards to the particle $i=1$. At the time $t_0 = 1/2v$, the moving particle collides with the first particle $(i=1)$.

Suppose that the collision is absolutely elastic. and therefore, energy and momentum are conserved. Then the zero particle stops at the point $x_1= 1/2^{1}$, and the first moves at a speed $v$ to the second particle located at  $x_2= 1/2^{2} =1/4$. The flying particle again will stop at this point, and the reflected one will move towards to the next particle. Consider what happens with the  $i$-th particle. This particle passes the interval $x_i -x_{i+1}=1/2^i -1/2^{i+1}=1/2^{i+1}$  in time $t_i =1/2^{i+1} v$. After this time, the  $i$-th particle stops. It’s easy to calculate the time all particles will collide
\[t_c =\sum_{i=1}^{\infty}\frac{1}{v 2^{i}} = \frac{1}{2 v}\]
In other words, after this finite time all particles will be in state of rest\footnote{If you doubt it, then try to indicate which particle is moving.}  This means that the initial kinetic energy $m v^2 /2$  of system disappeared in the process of an infinite number of absolutely elastic collisions. It might seem that the problem is the assumption of point size of particle, but it is not \cite{zen3}. It should be noted that the problem is even deeper. In such a process the principle of determinism of classical mechanics is violated. By the final state it is impossible even to find out whether any of the participants in the infinite series of particles moved before. Perhaps this gives an additional natural mechanism for the appearance of irreversibility in statistical physics. It is interesting to note that quantum mechanics is more loyal to determinism than classical mechanic.

Performing an infinite number of operations in a finite time creates interesting problems in purely mathematical fields. Recall that Alan Turing created theory of algorithms for solving the tenth Hilbert problem (find an algorithm to determine whether a given polynomial Diophantine equation with integer coefficients has an integer solution). He understood the microscopic structure of the algorithm and created a Turing machine (TM \cite{zen4,zen5}). Given Church's thesis, we can say that all algorithms correspond to TM. Of course, this thesis is not of a mathematical nature. It is based on observing that all the algorithms known so far can be represented as TM. Turing then proved the existence of mathematical problems that are algorithmically unsolvable. One of these problems is the problem of stopping a TM (see, for example \cite{zen6}). The problem of stopping. TM is quite simple. Is there an algorithm that by the "appearance" of TM and the initial data can say she will stop or not? \cite{zen6}). At first glance, the problem of stopping the TM does not look important. However, it is not so. If the problem of stopping TM had an algorithmic solution, then the fate of mathematics would be even more than sad.

In this case, a universal method of solving any mathematical problems would arise. Indeed, suppose you want to prove Great Fermi Theorem. This is a statement about the existence of integers x, y, z that satisfy the relation $x^n + y^n = z^n$ for $n> 2$. Then we construct TM, which performs the substitution of integers into this relation and stops if it is satisfied. But then there is a universal MT, which can answer the question whether this MT will stop and therefore to prove or to disprove Great Fermi Theorem as well as any other theorem.

G. Weil remembered Zeno \cite{zen7} when he proposed MT, which can be called Zeno's machine. This machine differed from MT only in the execution time of the $i$-th step. This machine carries out the $i$-th step in time  $2^{-i}$. Thus, Zeno machine can perform an infinite (countable) number of operations in a finite time. It is easy to prove that Zeno machine can solve the problem of stopping TM. Therefore, the reasons for which most mathematicians do not consider Zeno machines are understandable. A more detailed discussion of Zeno machines can be found, for example, in \cite{zen8,zen9}. It is funny to note that the problem of stopping the Zeno machine is not solvable on Zeno machine. At present, the view on the implementation of the laws of nature as on processing of information is gaining popularity. Therefore, one should expect attempts to use Zeno machines in this area.

A new outburst of interest in Zeno's paradox is associated with its quantum formulation in 1954  by Alan Turing. After \cite{zen10}, this formulation was called Zeno effect. Such $t$ leads, within the framework of quantum mechanics, to the fact that Achilles never really catches up with a turtle if the turtle carefully monitors Achilles behavior. Let's discuss this effect in details. Suppose that  the turtle checks the position of Achilles from the very start of the race. "Checks" in quantum mechanics means measuring the position of Achilles. This is not a trivial operation. Let the initial state of Achilles be  $|0\rangle$. In a short time $\Delta t$ according to the Schr\"{o}dinger equation his state will go into the state  $| \Delta t \rangle$ equal to
\[| \Delta t \rangle =e^{-\frac{i}{\hbar}H \Delta t} | 0\rangle \approx \left(1  - \frac{i}{\hbar}H \Delta t  - \frac{1}{2 \hbar^2}H^2 \Delta t^2 \right)| 0\rangle\]
where $H$  is a time-independent Hamiltonian of Achilles. The probability of finding Achilles in the initial state it is easily to evaluate
\[w=| \langle 0| \Delta t \rangle|^2 \approx 1- \frac{\Delta E^2}{\hbar^2}{\Delta t}^2\]
Here $\Delta E^2 =\langle 0|H^2 | 0 \rangle -\langle 0|H | 0 \rangle^2$ is the dispersion of energy and  $\frac{\Delta E}{\hbar}{\Delta t} \ll 1$. It naturally follows  from the relation obtained that this probability is close to 1. Now let the tortoise measures Achilles position n times at $\Delta t$ intervals. Then, due to the independence of measurements, the probability of finding Achilles at the starting point at a time $t=n \Delta t$  is determined by the product of the probabilities
\[ W_s = \left(1- \frac{\Delta E^2}{\hbar^2}{\Delta t}^2 \right)^n= \left(1- \frac{\Delta E^2}{\hbar^2}\frac{t^2}{n^2} \right)^n\]
Using $\lim_{n \to \infty} (1+1/n)^n \to e$  easy to get
\[W_s = \left(1- \frac{\Delta E^2}{\hbar^2}\frac{t^2}{n^2} \right)^n \xrightarrow[n\to \infty]{} e^{- \frac{\Delta E^2}{\hbar^2}\frac{t^2}{n}} \xrightarrow[n\to \infty]{}  1 \]
In other words, if the turtle really wants to win the race, then it must watch Achilles with infinite frequency. This guarantees it (the turtle) that he (Achilles) gets is completely stuck at the start. This is the quantum Zeno effect. The physical meaning of this effect is related with reduction of wave function during the process of measurement. We can show this using simplified considerations. Indeed, the state of Achilles after a short period of time can be considered as a superposition of his the initial state $|0\rangle$  and the state of the shifted Achilles  $| 1 \rangle$, i.e.  $|\Delta t\rangle =\alpha |0\rangle + \beta | 1 \rangle$, where $\alpha$ and $\beta$ are complex amplitudes, that determine the probability of detecting corresponding state during the process of measurement. Considering that time is short, the initial state will be obtained as a result of measurement with a greater degree of probability. In the process of measuring state $|\Delta t\rangle =\alpha |0\rangle + \beta | 1 \rangle$ transforms into state $|0\rangle$ . After all, this initial state begins to evolve another time but the next measurement returns Achilles once again to its original position. Of course, these simplified considerations can be confirmed by more detailed calculations (see, for example, \cite{zen10,zen11}). But despite technical difficulties, Zeno effect was realized even experimentally \cite{zen12}.

What is the secret of such incessant interest in Zeno's paradoxes? Apparently, this is due to the daunting task that Zeno set himself. His goal is to in an attempt to logically prove that of the two ideas of discreteness or continuity space and time, only one is logically consistent. This problem remains relevant even now. Moreover, the choice between these ideas follows from the need to combine GR and quantum theory. These theories are based on opposite ideas about the nature of things. Modern science gives preference for continuity by acquiring a powerful mathematical apparatus such as mathematical analysis or relatively recent used non-standard analysis \cite{zen13}.

The fight of these two concepts has already played an extremely important role in creating modern mathematics. However, the Zeno goal  was not achievable. Both of these concepts are logically consistent. Consequently, the issue of the applicability of these ideas to the physical reality cannot be defined by logic or mathematics.

There are physical considerations for which the continuum model is a nonphysical idealization and that in any finite volume can be found only by a finite number of degrees of freedom. This considerations include the existence of a minimal Planck length as well as black holes. In the following sections, we will touch on some problems associated with this.

\section{Quantum manifestations of classical gravity}

In this section we consider the simplest consequences of inclusion in quantum reality gravitational effects. In the absence of real experiments, we will rely on the arguments obtained as a result of the analysis of thought experiment (Gedankenexperiment). The most important consequence of the consideration of gravitational effects is the inevitability of the transition from the classical concept of continuum to discrete space-time. Below we will analyze in detail the physical mechanisms associated with the introduction of minimum length.

\subsection{Generalized Uncertainty Principle}

Generalized Uncertainty Principle (GUP) is the basis of the phenomenology of Planck scale physics. The need to generalize the original Heisenberg Uncertainty Principle (HUP) is associated with the following obvious reason: HUP is incompatible with GR. The fact is that any measurement process is accompanied by the transfer of energy. This energy distorts background space-time, introducing additional uncertainty into the position of the measured object. GUP eliminates this inconsistency.

Initially GUP was derived in string theory and in a number of generalizations of this theory \cite{24s,25s,26s,27s}. These works raised the following question: can this principle, due to its simplicity and generality, be obtained outside the framework of a specific quantum theory? An affirmative answer to this question was received very soon. Let’s consider the simplest conclusion of the GUP \cite{28s,29s}. More rigorous conclusions do not change the final result.

Using a photon with a wavelength $\lambda$, we can localize the position of the object, for example, an electron, with accuracy
\begin{equation}\label{e2_1}
  \Delta {x_e} \ge \lambda
\end{equation}
Since  $\lambda  = \hbar /p$, where $p$ is the photon momentum and the maximum value of  $\Delta p \approx p$, then
\begin{equation}\label{e2_2}
  \Delta {x_e} \ge \frac{\hbar }{{\Delta p}}
\end{equation}
The gravitational interaction between the photon and the electron was not taken into account in this expression.

Let us now evaluate the gravitational interaction of a photon with an electron in the rough Newtonian approximation, assuming that the photon behaves like a classical particle with effective mass $E/{c^2}$, where $E$ is the photon energy. Let the interaction take place in areas of characteristic size $L$. Acceleration of an electron caused by gravitational interaction with a photon
\begin{equation}\label{e2_3}
  \left| {\ddot x} \right| = \frac{{GE/{c^2}}}{{{x^2}}}
\end{equation}
where $x$ is the distance between electron and photon. For a typical interaction time $L/c$ electron displacement is
\begin{equation}\label{e2_4}
  \Delta {x_g} \approx \frac{{GE}}{{{c^2}{x^2}}}{\left( {L/c} \right)^2}
\end{equation}
For $x \approx L$ and $E = pc$, we find
\begin{equation}\label{e2_5}
  \Delta {x_g} \approx \frac{{Gp}}{{{c^3}}}
\end{equation}
Given that the uncertainty of the electron momentum should be on the order of the photon momentum and introducing the Planck length ${l_{Pl}} = \sqrt {\frac{{\hbar G}}{{{c^3}}}}$ , we get
\begin{equation}\label{e2_6}
  \Delta {x_g} \approx l_{Pl}^2\frac{{\Delta p}}{\hbar }
\end{equation}
Summing up the contributions of electromagnetic and gravitational interactions, we finally obtain
\begin{equation}\label{e2_7}
  \Delta x \ge \frac{\hbar }{{\Delta p}} + l_{Pl}^2\frac{{\Delta p}}{\hbar }
\end{equation}
HUP can be considered as a limiting case of GUP
\begin{equation}\label{e2_8}
  \left( {GUP,\;\Delta p \to 0} \right) \to \left( {HUP} \right)
\end{equation}
Let's transform the ratio (\ref{e2_7}) to the form
\[ \frac{\Delta x}{c} \ge \frac{\hbar }{{\Delta p}c} + l_{Pl}^2\frac{{\Delta p c}}{\hbar c^{2}}\]
Introducing uncertainties of time $\Delta t = \frac{\Delta x}{c}$  and energy $\Delta E = {\Delta p}c$ we obtain
\begin{equation}\label{e2_8a}
   \Delta t \ge \frac{\hbar }{\Delta E} + t_{Pl}^2\frac{{\Delta E}}{\hbar }
\end{equation}
This relationship can be interpreted as a generalization that takes into account gravitational effects, of energy-time uncertainty relationship. It is easy to see that as $\Delta E \to 0$ this inequality becomes the usual inequality $\Delta t \Delta E \geq \hbar$.
We reproduce the result (\ref{e2_6}) using the field equations
\begin{equation}\label{e2_9}
  {R_{\mu \nu }} - \frac{1}{2}{g_{\mu \nu }}R = \frac{{8\pi G}}{{{c^4}}}{T_{\mu \nu }}
\end{equation}
The left side of (\ref{e2_9}) is of the order of  $\frac{{\delta g}}{{{L^2}}}$, and the right - $\frac{{8\pi GE}}{{{c^4}{L^3}}} \approx \frac{{Gp}}{{{c^3}{L^3}}}$, here $L$ is a characteristic linear size. Hence,
\begin{equation}\label{e2_10}
  \delta g \approx \frac{{Gp}}{{{c^3}L}}
\end{equation}
Uncertainty in the measurement of distances leads to fluctuations in the metric
\begin{equation}\label{e2_11}
  \delta g \approx \frac{{\Delta x}}{L}
\end{equation}
Comparing (\ref{e2_10}) and (\ref{e2_11}), we return to the result (\ref{e2_6}).

Modification of HUP taking into account gravitational effects can be presented in the form
\begin{equation}\label{e2_12}
  \Delta x\Delta p \ge \frac{\hbar }{2}[1 + {\beta _0}{\left( {\Delta p/{m_{Pl}}c} \right)^2}]
\end{equation}
Here  ${\beta _0}$ is a free parameter. Recall that the HUP is consequence of the non-commutativity of the corresponding operators of the observed quantities.

Therefore, we can assume that this modification is a consequence of the deformation of commutation relations
\[\Delta x \Delta p \geq \frac{1}{2} \langle [x, p ] \rangle\]
\begin{equation}\label{e2_13}
  [x, p] =i \hbar (1+ \beta_0 (\Delta p / m_R c)^2)
\end{equation}
Current measurement accuracy limits  ${\beta _0} < {10^{33}}$. This excludes the existence of an intermediate length scale up to ${10^{ - 19}} \, m$. Modification, on the Planck scale, corresponds to   ${\beta _0} \approx 1$.

\begin{figure}
  \centering
    \includegraphics[width=6 cm]{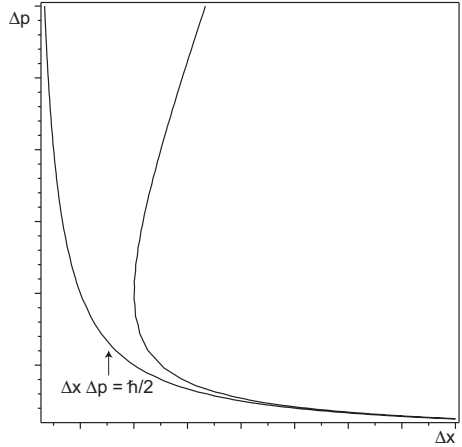}\\
  \caption{The boundary of the region allowed in the process of measuring оf the position of the particle, taking into account gravitational interaction and without it.}\label{fg1}
\end{figure}

The dramatic consequence of taking gravity into account in the measurement process is the emergence of such a fundamental concept as the minimum length. The physics of this parameter is extremely simple. At small photon momenta (large wavelengths), the localization of the measured object will be bad. For large photon momenta, its gravitational interaction with the measured object will again negatively affect the process of measurements. Between these two extremes, we can choose the photon momentum, which optimizes the measurement process. Figure \ref{fg1} shows the boundaries of the regions admissible in the process of measuring the position of a particle, taking into account gravity (GUP) and without taking into account gravity (HUP).

\subsection{Minimum length}

Probably the most important consequence of the transition from HUP  to GUP is the emergence of a minimum spatial length. Indeed, minimizing relation (\ref{e2_7}) we find
\begin{equation}\label{e2_14}
  \Delta {x_{\min }} = 2\sqrt {\frac{{G\hbar }}{{{c^3}}}}  = 2{l_{Pl}}
\end{equation}
Minimum realized for
\begin{equation}\label{e2_15}
  \Delta {p_{\max }} = \sqrt {\frac{{\hbar {c^3}}}{G}}  = {m_{Pl}}c
\end{equation}
What is the physical mechanism for generating the minimum length? How is it related to gravity?

In exactly the same way, inequality (\ref{e2_8a}) implies the minimal time scale
\begin{equation}\label{e2_15a}
    \Delta t_{min}=2 t_{Pl}
\end{equation}
which is achieved with a maximal energy uncertainty equal to
\[\Delta E_{max} =\frac{\hbar}{t_{Pl}}\]
Thus, the minimum time interval is also finite and is determined by the Planck time scale.

The appearance of singularities in the theory is considered to be the first signal that the theory has gone beyond its applicability and needs to be modernized. Natural  variant of such modernization consists in taking into account effects that have fallen from consideration. This allows us to hope to make the theory free of divergences. There is an alternative point of view formulated by R. Penrose \cite{30s} in the form of cosmic censorship hypothesis (principle). According to this hypothesis. the singularities of space-time appear in places that, like the inner regions of the black holes, are hidden from the observer. A natural question arises: is it possible to generalize cosmic censorship principle to the level of the physic censorship principle, making it an universal physical principle? Existence of minima length allows you to give an affirmative answer to this question.  The formation of the event horizon is the mechanism for generating of the minimum length. In the simplest case of Schwarzschild solution to Einstein's field equations the event horizon is a hypothetical sphere around the gravitational point singularities dividing the space into two causally unrelated areas. The gravitational radius ${r_g}$ is the radius of this sphere for a body of mass $m$
\begin{equation}\label{e2_16}
  {r_g} = \frac{{2mG}}{{{c^2}}}
\end{equation}

Let us now consider how this mechanism works on the Planck scale. First we show that for a particle with a mass of the order of Planck’s, the gravitational radius coincides with the Compton wavelength $\lambda_c$. Indeed,
\[{\lambda _c} = {r_g}\; \to \frac{\hbar }{{mc}} = \frac{{2mG}}{{{c^2}}},\]
\begin{equation}\label{e2_17}
  m = \sqrt {\frac{{\hbar c}}{{2G}}} ,\quad {m_{Pl}} = \sqrt {\frac{{\hbar c}}{G}}
\end{equation}
As we saw above (see relations (\ref{e2_9})-(\ref{e2_11}))
\begin{equation}\label{e2_18}
  \frac{{\delta g}}{{{L^2}}} \approx \frac{{GE}}{{{c^4}{L^3}}}
\end{equation}
Localization of energy should not contradict quantum mechanics,
\begin{equation}\label{e2_19}
  \frac{L}{c} \approx \frac{\hbar }{E}
\end{equation}
Substituting $L \approx \frac{{\hbar c}}{E}$ in relation (\ref{e2_18}) we find
\begin{equation}\label{e2_20}
  \delta g \approx \frac{{G{E^2}}}{{{c^5}\hbar }}
\end{equation}
Gravitational perturbation is significant if $\delta g \approx 1$ and, therefore, $E = {m_{Pl}}{c^2}$. But, as we saw above (\ref{e2_17}), just such a perturbation of space-time will lead to the formation of an event horizon.

Discrete space-time is a very old idea, not directly related to gravity. It is now becoming clear that M. Bronstein was the first to find a connection between these  concepts. Already in 1936 he claimed \cite{31s} that gravity is different from others fundamental interactions, since gravity does not allow an arbitrarily large concentration of energy in a finite region of space.Bronstein paid attention to the fact that, since the gravitational radius should be less than the linear dimensions of the body, the possibility of measurements in extremely small areas is limited. Therefore, without changing the quantum-mechanical commutation relations, it is hardly possible to combine quantum theory and gravity. Thus, more than 80 years ago, Bronstein formulated an almost modern concept of the role of gravity in the process of precision measurements.

\subsection{Principle of maximal force}

The role of the event horizon as the boundary of the observed physical reality becomes more understandable, if we turn to the principle of maximal force. The statement about the existence within the framework of GR of maximal (limit) force  as a principle was first formulated by G. Gibbons \cite{22s}. This force is equal
\begin{equation}\label{e2_21}
  {F_{\max }} = \frac{{{c^4}}}{{4G}} \approx 3.25 \times {10^{43}}\;N
\end{equation}
This limit does not depend on nature of the forces and is valid for gravitational, electromagnetic, nuclear or any other forces. It is interesting to note that up to a factor of $2 \pi$ the reciprocal of the maximal force coincides with the energy-momentum tensor in field equations of GR.

Absolutely equivalent statement is the existence of the maximal power
\begin{equation}\label{e2_22}
  {P_{\max }} = \frac{{{c^5}}}{{4G}} \approx 9.07 \times {10^{51}}W
\end{equation}
Maximum force and power are components of a 4-vector $\frac{{d{p^\lambda }}}{{dt}}$, where $p_i$ is 4-momentum.  Maximum force and power are invariants, which follows from invariance of the quantities c and $G$. Its time dependence however is not excluded. The limit holds both for each component of the 3-force, and for its absolute value.

Maximal power allows a trivial physical interpretation. Consider the power released during the "annihilation" of a black hole of mass $m$. The minimal the time for the realization of such a process is the time that light passes through its gravitational diameter  $t = 2{r_g}/c = \frac{{4mG}}{{{c^3}}}$,
\begin{equation}\label{e2_23}
  P = \frac{{m{c^2}}}{{4mG/{c^3}}} = \frac{{{c^5}}}{{4G}},
\end{equation}
which exactly coincides with the limit power introduced above.

The limit power is the long-lived GR. Gravitational luminosity (total power lost on gravitational radiation)
\begin{equation}\label{e2_24}
  L_{GR}=\frac{G}{5c^5} \langle \dddot{Q}_{ij} Q^{ij} \rangle \to L_{GR}^{max}=\frac{c^5}{G}
\end{equation}
Here $Q$ is the mass quadrupole moment of the system.  The upper limit of luminosity in nature, the so-called natural luminosity, introduced by Einstein, up to the numerical factor coincides with the limit power.

Let us now dwell on a fundamentally important issue - the mechanism of the appearance of maximal force \cite{33s}. Consider two bodies with masses $m_1$ and $m_2$ separated  by the distance $R$. In Newtonian mechanics
\begin{equation}\label{e2_25}
  F = G\frac{{{m_1}{m_2}}}{{{R^2}}} = \left( {\frac{{G{m_1}}}{{{c^2}R}}} \right)\left( {\frac{{G{m_2}}}{{{c^2}R}}} \right)\frac{{{c^4}}}{G}
\end{equation}
since  ${m_1}{m_2} \le \frac{1}{4}{({m_1} + {m_2})^2}$, then
\begin{equation}\label{e2_26}
  F \le {\left[ {\frac{{\left( {{m_1} + {m_2}} \right)G}}{{{c^2}R}}} \right]^2}\frac{{{c^4}}}{{4G}}
\end{equation}
The approach of bodies is limited by the condition  $R > {r_g}$, which prevents the formation of black hole with mass $m_1 + m_2$. Consequently
\begin{equation}\label{e2_27}
  F \le \frac{{{c^4}}}{{4G}}
\end{equation}
Surfaces on which the maximal force (maximal momentum flux) or maximal power (maximal energy flux) is realized represent the event horizon. Any attempt to exceed the force limit results in a horizon, which prevents a further increase in force.

It is important to note that the maximal force cannot be realized in volume. If this were possible, then by performing the Lorentz transformation it is possible to move on to a larger value of force. Thus, the maximal force can be realized only on surface, but not in volume. In addition, these surfaces must be unattainable within the framework of GR. Such surfaces represent event horizons. Maximal force and maximal power are achieved only at the horizons.

The relationship between the concepts of horizon and limit force plays a central role in obtaining the equations of GR.  Just as in the special theory of relativity, the Lorentz transformations are a consequence of the existence of a limit velocity, the presence of a limit force leads to field equations of GR \cite{20s,21s}.

In view of the fundamental importance of this result, we give one more example, clarifying mechanism for the connection of limit force and minimal length. In Newtonian mechanics $F = dp/dt$ and therefore
\begin{equation}\label{e2_28}
  {F_{max}} = \frac{{{{(\Delta p)}_{\max }}}}{{{{\left( {\Delta t} \right)}_{\min }}}} \approx \frac{{mc}}{{{t_{Pl}}}} = \frac{{m{c^2}}}{{{l_{Pl}}}}
\end{equation}

At first glance, it seems that by unlimitedly increasing the mass, we can get arbitrarily great force. However, this is not. Restriction is associated with the occurrence horizon at a fixed length scale  $\left( {{l_{Pl}}} \right)$. Indeed,  omitting the numerical factors $O(1)$, we find the mass for which the gravitational radius equal to the Planck length
\begin{equation}\label{e2_29}
  m \approx \frac{{{l_{Pl}}{c^2}}}{G} = \sqrt {\frac{{\hbar G}}{{{c^3}}}} \frac{{{c^2}}}{G} = \sqrt {\frac{{\hbar c}}{G}}  = {m_{Pl}}
\end{equation}
Consequently, the maximum mass that we can use in (33)and which prevents the formation of the horizon is the Planck mass, and therefore
\begin{equation}\label{e2_30}
  {F_{\max }} \approx \frac{{{m_{Pl}}{c^2}}}{{{l_{Pl}}}} = \frac{{{c^4}}}{G}
\end{equation}
Obviously, the result (35) can be obtained from dimensional considerations
\begin{equation}\label{e2_31}
  {F_{Pl}} = {m_{Pl}}\frac{{{l_{Pl}}}}{{t_{Pl}^2}} = \sqrt {\frac{{\hbar c}}{G}} \sqrt {\frac{{\hbar G}}{{{c^3}}}} \frac{{{c^5}}}{{\hbar G}} = \frac{{{c^4}}}{G}
\end{equation}
Note that all our statements apply only to $D = 3 + 1 = 4$.

Consider another important gedanken experiment that allows us to understand the role of quantum effects in the formation of an event horizon \cite{28s}. This experiment allows to detect a new mechanism of violation of the particle localization measure, compared with HUP. The experiment is to the maximum possible compression of the volume containing the mass $m$. Compression is assumed isotropic. By the maximum possible, we mean compression to the Schwarzschill sphere of radius ${r_g} = 2mG/{c^2}$.  With further compression, a black hole is formed. Of course gravitational radius ${r_g} = 2mG/{c^2\to 0}$ as $m \to 0$.

From the HUP it follows that the uncertainty in momentum in a volume with a characteristic size $l$  at least order $\Delta p \approx \hbar /l$. Because the energy in the volume is given by the relation ${E^2} = {M^2}{c^4} + {p^2}{c^2}$ energy uncertainty $\Delta E \approx c\Delta p \approx \hbar c/l$. When the volume is compressed, we can reach such small values that this energy uncertainty increases to $\approx 2m{c^2}$ after which it will become possible particle-antiparticle pair production. In this case, localization is destroyed, and further volume reduction does not make sense. This limit volume has energy and size.
\[m{c^2} \approx \Delta E \approx \hbar c/l_c ,\;\;l_c \approx \hbar /mc\] 	
The length $l_c$  represents the limit of localization in quantum mechanics. Therefore we have two characteristic sizes describing the process of compression:  $r_g$ and  $l_c$. As we saw above for $m = {m_{Pl}}$, these values coincide and  $l_c = {l_{Pl}}$.  Thus, we once again became convinced that the combination of gravity and quantum effects creates an insurmountable obstacle to further compression on the linear sizes  $l \approx {l_{Pl}}$.

\subsection{Modified Planck units}

As noted earlier, introduction of the Planck system of units] prior to creation of quantum mechanics as well as of special and general relativity was a significant event in physics which well ahead of time. Even after more than a century, many questions remain, the answers to which remain not found. The depth of these issues may become clear only after creating a quantum theory of gravity. The Planck units can now be regarded as a kind of laboratory that allows b to look into the future. From this point of view they are of great interest.

It should be noted that the choice of the initial constants necessary for constructing the fundamental scales, $\hbar, c, G$ is not unique.

The various combinations of these constants can be used as starting material. There are many examples where  using of a particular combination of fundamental constants greatly simplifies the theory. A classic example is the fine-structure constant  $\alpha  = {e^2}/\hbar c$.  It is difficult to imagine quantum electrodynamics without using of this combination.

Discussion of various issues related to Planck units system a huge number of publications have been devoted (see, for example,\cite{34s}), affecting both physics elementary particles and cosmology. Among such works, it should be noted attempts to construct alternative sets of fundamental scales, differing both in the choice of fundamental constants and their number \cite{35s,36s}. It seems natural to modify the Planck unit system ones in such a way that only limit values enter into it \cite{37s,38s}. For this, the gravitational constant $G$  should be excluded from the initial set $( \hbar , c, G )$. This constant is not a limit value and it should be expressed in terms of limit power
\[\eta  \equiv {P_{max}} = \frac{{{c^5}}}{G}\]
or minimal length  ${l_{\min }} = {l_{Pl}}$. (We omit factors of $O(1)$.)

In the first case, by replacement $\left( {\hbar ,c,G} \right) \to \left( {\hbar ,c,\eta } \right)$ we obtain a modified Planck units system
\[{m_{Pl}} = \sqrt {\frac{{\hbar \eta }}{{{c^4}}}} ,\]
\[{l_{Pl}} = \sqrt {\frac{{\hbar {c^2}}}{\eta }}\]
\begin{equation}\label{e2_32}
  {t_{Pl}} = \sqrt {\hbar /\eta }
\end{equation}

and in the second  case, by replacement $\left( {\hbar ,c,G} \right) \to \left( {\hbar ,c,{l_{\min }} = {l_{Pl}}} \right)$
\[{m_{Pl}} = \frac{\hbar }{{{l_{Pl}}c}},\]
\[{l_{Pl}},\]
\begin{equation}\label{e2_33}
  {t_{Pl}} = \frac{{{l_{Pl}}}}{c}
\end{equation}
The limit values $\eta$ and $l_{Pl}$  are are interconnected
\begin{equation}\label{e2_34}
  \eta  = \frac{{{c^5}}}{G} = \frac{\hbar }{{t_{Pl}^2}} = \frac{{\hbar {c^2}}}{{l_{Pl}^2}}
\end{equation}
It should be emphasized again that the necessary condition for the existence of the event horizon is finiteness of the realized power and of the speed of light. Thereat, as we have already noted, the magnitude of the limit value is less significant than the fact of its existence. It is easily seen that
\[\mathop {\lim }\limits_{c \to \infty } {r_g} = \mathop {\lim }\limits_{c \to \infty } \frac{{2mG}}{{{c^2}}} = 0;\]
\begin{equation}\label{e2_35}
  \mathop {\lim }\limits_{{P_{\max }} \to \infty } {r_g} = \mathop {\lim }\limits_{c \to \infty } \frac{{2m{c^3}}}{{{P_{\max }}}} = 0
\end{equation}
In other words, $\eta  \to \infty $ or $c \to \infty $, the concept of the gravitational radius and, therefore, the event horizon loses its meaning. Choice as a new fundamental constant maximal power or minimal length leads to modified Planck units, keeping the previous numerical values. However, this transition opens up interesting possibilities both for the interpretation of already known results and for obtaining new results.

\subsection{Space-time foam}

Space-time in GR is perceived as an unconditionally classical object. An attempt to study its properties on a small scale inevitably leads to the necessity of attracting quantum theory. If a space undergoes quantum ?uctuations, the latter must manifest themselves as uncertainties in different kinds of measurements. Apparently, J. Wheeler \cite{39s} was the first to appreciate the importance of quantum fluctuations. After that, the image of the fluctuating, foamy space-time entered as a fundamental element of the physical picture of the world.

\begin{figure}
  \centering
  \includegraphics[width=7 cm]{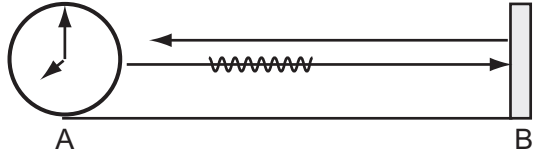}\\
  \caption{Wigner gedanken experiment.}\label{fg2}
\end{figure}

Length measurement is one of the most important types of measurements. Using simple considerations, Wigner \cite{40s} considered the problem of the uncertainty  in measuring the distance between two points. For this, he suggested gedanken thought experiment (see Fig.\ref{fg2})

Let the clock be at the beginning of the measured segment, and the mirror at the end. By sending a photon to a distant point and recording the return time, we can determine the distance between the points. Let there be an uncertainty $\delta l$ in the position of the clock associated with the presence of quantum fluctuations. Then the total uncertainty in the calculated position of the clock consists of two contributions
\[\Delta l = \delta l + \frac{{\delta p}}{m} \cdot \frac{l}{c} \]
Here $\Delta l$ is the total uncertainty in the distance, $\delta p$ is the uncertainty in the clock momentum when the signal is received, $l$  is distance between points, $m$ is mass of clock. It is easy to evaluate the uncertainty in the clock momentum, given the uncertainty in the wavelength
\[\delta p = \frac{\hbar }{{\delta l}}\]
After substitution we get
\[\Delta l = \delta l + \frac{{\hbar l}}{{\delta l \cdot mc}}\]
It remains only to calculate the value of $\delta l$ at which $\Delta l$ is minimal. Then
\[\delta {l_{\min }} = \sqrt {\frac{{\hbar l}}{{mc}}} \]
Consequently,
\begin{equation}\label{e2_36}
  \delta {l^2} \ge \frac{{\hbar l}}{{mc}}
\end{equation}
At first glance, it seems that the uncertainty in measuring distance can be reduced unlimitedly, increasing the mass of clock. So, as $m \to \infty $, uncertainty of distance measurement due to quantum fluctuations $\delta l \to 0$. However the clocks not are an abstract object: they obey physical laws. These laws include the laws of quantum physics and general relativity. It is automatic means that the clock must satisfy additional restrictions. So size of clock  $d$  must exceed their gravitational radius
\begin{equation}\label{e2_37}
   d > \frac{{Gm}}{{{c^2}}}
\end{equation}
Otherwise the watch will turn into a black hole and its readings will become inaccessible to us. In addition, for the implementation of a gedanken experiment it is necessary to fulfill the conditions  $\delta l \geq d$. Then, using inequalities (\ref{e2_36}) and (\ref{e2_37}), we obtain \cite{41s}
\begin{equation}\label{e2_38}
  \delta l \ge {\left( {ll_{Pl}^2} \right)^{1/3}} = {l_{Pl}}{\left( {\frac{l}{{{l_{Pl}}}}} \right)^{1/3}},\quad {l_{Pl}} \equiv {\left( {\frac{{G\hbar }}{{{c^3}}}} \right)^{`/2}}
\end{equation}
The resulting inequality defines a fundamental limitation on the accuracy of distance measurement, regardless of the measurement method used. Similarly, you can associate the minimum uncertainty in measuring time with a measured time interval. Of course, in this case, we are not talking about the accuracy of a particular clock  design, but about universal limitations on the accuracy of time measurements. Such restrictions are based on fundamental physical laws. To establish the uncertainty of time measurements, we again turn to the gedanken experiment. Relations (\ref{e2_36}) and (\ref{e2_37}) can be written in terms of uncertainty $\delta t = \delta l/c$  when measuring time $t = l/c$,
\begin{equation}\label{e2_39}
  \delta {t^2} \ge \frac{{\hbar t}}{{m{c^2}}},\;\;\delta t \ge \frac{{Gm}}{{{c^3}}}
\end{equation}
Combining these inequalities, we obtain
\begin{equation}\label{e2_40}
  \delta t \ge {\left( {tt_{Pl}^2} \right)^{1/3}}
\end{equation}
Relation (\ref{e2_40}) sets the minimum uncertainty when measuring an arbitrary time interval $t$. In order to estimate the scale of such uncertainty, let's use the age of the Universe as the maximum time interval. In this case $\delta t \sim {10^{ - 23}}$ sec. It should be noted that the absolute value of the uncertainty grows   while relative   decreases. Inequality (\ref{e2_40}) can be given another meaning. To do this, we resolve it with respect to the measured time interval $t$
\begin{equation}\label{e2_41}
  t \le \delta t{\left( {\frac{{\delta t}}{{{t_{Pl}}}}} \right)^2}
\end{equation}
This inequality can be understood as a restriction on the measured time interval $t$. However, in meaning, this interval corresponds to the time of the correct functioning of the clock or, in other words, "the time of their life" \cite{43s}. Therefore, we come to the surprising conclusion that the more accurate the clock, the less time it functions with a given accuracy. What can we count on when buying clock, say, femtosecond accuracy? Using $\delta t = {10^{- 15}}$ seconds from inequality (\ref{e2_41}), we find
\begin{equation}\label{e2_42}
  t \le {10^{34}}\;y
\end{equation}
Thus, if we could buy them at the time of the birth of the universe they could function much longer than the current age of the universe.
Inequality (\ref{e2_41}) leads to one interesting and important consequence. Let's use black hole as a clock. Then, as the accuracy $\delta t$ of such a "device", it is natural to choose $\delta t = {r_g}/c$. Now, using inequality (\ref{e2_40}), we determine the lifetime of such an "alarm clock" \cite{42s}
\begin{equation}\label{e2_43}
  t \le \frac{{{m^3}{G^2}}}{{\hbar {c^4}}}
\end{equation}
The obtained lifetime of the black hole in such an unusual way coincides with the time of evaporation of the black hole, which Hawking calculated. The difference is only in numerical multiplier.

The uncertainty in measuring lengths and times automatically leads to the uncertainties of the metric tensor,
\begin{equation}\label{e2_44}
  \delta {g_{\mu \nu }} = \left( {{{\left( {\frac{{{l_{Pl}}}}{l}} \right)}^{2/3}},{{\left( {\frac{{{t_{Pl}}}}{t}} \right)}^{2/3}}} \right)
\end{equation}
The clocks appearing in the previous arguments are closely related to special relativity and were used in it for conducting gedanken experiments. It is interesting to find out what features in the measurement of space-time intervals will arise using the elements of  quantum physics. The formalism based on GUP allows us, in particular, to answer the question, why is it impossible to construct ideal quantum clocks \cite{42s,44s}? For this, in a gedanken experiment we use a different type of clock. As a concrete implementation of the clock, we consider quantum clocks based on observation of radioactive decay, which is described by the equation
\begin{equation}\label{e2_45}
  \frac{{dN}}{{dt}} =  - \lambda N
\end{equation}
where $N (t)$ is the current number of radioactive particles in the sample. The average number of decayed particles (nuclei) during the time  $\Delta t \ll {\lambda ^{ - 1}}$ is $\Delta N = \lambda N\Delta t$, which makes it possible to measure the time intervals by counting the number of decaying particles:
\begin{equation}\label{e2_46}
  \Delta t = \frac{{\Delta N}}{{\lambda N}}
\end{equation}
The relative error of this method of  time measurement  $\varepsilon  = {\left( {\lambda N\Delta t} \right)^{ - 1/2}} = 1/\sqrt {\Delta N}  \le 1$.  It would seem that by increasing the number N of  decays, it is possible with the help of such a process unlimited increase accuracy of time intervals measurement. Increase in the number of decays inevitably associated with increase in the mass (size) of the clock. However, such a process is limited a condition: an increase in mass should not lead to the transformation of a watch into black hole (i.e., horizon formation). Let us analyze what quantitative restrictions this condition will to lead. Using the uncertainty principle $\Delta E\Delta t \ge \hbar /2$, we can transform relation (\ref{e2_46}) into the inequality
\begin{equation}\label{e2_47}
  \Delta t \ge \frac{\hbar }{{2{\varepsilon ^2}{c^2}}}\frac{1}{M}
\end{equation}
where $M = N{m_p}$ is the mass of  the clocks (${m_p}$ is the mass of one particle). If the radius of the clock R (we assume that they have a spherical shape) becomes less than the gravitational radius $r_g$, we will lose the ability to use the clock to measure time. Condition $R > {r_g}$ transforms into
\begin{equation}\label{e2_48}
  \frac{1}{M} > \frac{{2G}}{{{c^2}R}}
\end{equation}
Substituting relation (\ref{e2_48}) into (\ref{e2_47}), we obtain
\begin{equation}\label{e2_49}
  \Delta tR > \frac{1}{{{\varepsilon ^2}}}\frac{G}{{{c^4}}}\hbar
\end{equation}
Understanding by $R$ the uncertainty $\Delta r$ in the position of a physical object, on the basis of which the process of  time measurement is constructed and taking into account that $\varepsilon  \le 1$, we finally find
\begin{equation}\label{e2_50}
  \Delta t\Delta r > \frac{G}{{{c^4}}}\hbar
\end{equation}
The resulting inequality limits the ability to determine the temporal and spatial coordinates of the event with arbitrary accuracy. Let us analyze expression (\ref{e2_50}) using the concept of maximal force  (\ref{e2_21}). For this, we represent (\ref{e2_50}) in the form
\begin{equation}\label{e2_51}
  \Delta t\Delta r > \frac{1}{{{F_{\max }}}}\hbar
\end{equation}

For a fixed Planck constant   $\hbar $, only the  maximal force  ${F_{\max }}$  determines the constraint superimposed on the size of a quantum clock. If in theory there is no limit force, i.e., ${F_{\max }} = \infty $, then ${r_g} \to 0$  and there is no restriction on the size of a quantum clock. The main reason for limiting the size of a quantum clock is the requirement  $R > {R_g}$, equivalent to preventing horizon formation. Therefore, the appearance in the relation (\ref{e2_5}) of the force ${F_{\max }}$, which can be achieved only on the horizon, seems absolutely natural. The structure of relation (\ref{e2_50}), in which there is no information about specific clock design  suggests that this relation can be obtained from fairly general considerations. Indeed , we use the uncertainty relation for this
\begin{equation}\label{e2_52}
  \Delta {x_{\min }}\Delta {p_{\max }} \ge \frac{\hbar }{2}
\end{equation}
Given that ${F_{\max }} = \frac{{\Delta {p_{\max }}}}{{\Delta {t_{\min }}}}$  we will immediately receive the minimum clock size   required for measuring time interval   obeys the constraint
\begin{equation}\label{e2_53}
  \Delta {x_{\min }}\Delta {t_{\min }} \ge \frac{\hbar }{{{F_{\max }}}} = \frac{{\hbar c}}{\eta }
\end{equation}
in full accordance with (\ref{e2_51}). Actually, this relation determines the structure of space-time foam! In other words, the cellular structure of the configuration space-time is determined by the relation quite similar to division quasiclassical phase space per cell, taking into account the HUP. The difference is in the value of the minimum volume, which for space-time is determined by three limit values. The simplicity of relation (\ref{e2_53}) emphasizes the fundamental nature of all three limit values  $\hbar$, $c$, $\eta $. Of course, the previously obtained restrictions on the limits of measurability of distances and time (\ref{e2_38}) and (\ref{e2_40}) are consistent with relations (\ref{e2_53}). Indeed, multiplying these uncertainties, we obtain
\begin{equation}\label{e2_54}
  \delta l \cdot \delta t \ge {\left( {\frac{{l\hbar }}{{c\eta }}} \right)^{1/3}}c{\left( {\frac{{t\hbar }}{\eta }} \right)^{1/3}} = {\left( {l \cdot t} \right)^{1/3}}{\left( {\frac{{\hbar c}}{\eta }} \right)^{2/3}}
\end{equation}
Suppose we measure the minimum spatial and time scales i.e. $l = \delta l$ and $t = \delta t$. In this case (\ref{e2_54}) reproduces the relation obtained above (\ref{e2_53}).

\section{New physics generated by the synthesis of quantum mechanics and gravity}

The existence of a minimum length and, as a consequence, the discrete structure of space on the one hand are, practically unrestricted source of new effects which are absent in the spatial continuum, and, on the other hand, allows you to detect previously unknown relationships between the phenomena already studied. At the same time, any theory with a minimum length inevitably contradicts a number of traditional formulations of both quantum mechanics and GR. In this section, we will consider a number of issues from to this area.

\subsection{The final stage of black holes evolution}

In the process of evolution, a dynamic system can go through various spatial and energy scales. An intriguing example of this situation is black hole evolution. This evolution occurs due to Hawking radiation. Concept of Hawking radiation is based on  a quantum vacuum. A quantum vacuum is filled with a mixture of particles and antiparticles, which are constantly created and destroyed. This quantum concept of vacuum follows the HUP, which allows quantum vacuum fluctuations. In a strong gravitational field in the vicinity of the black hole event horizon, a particle and an antiparticle may become separated by a distance of Compton wavelength $\lambda$ that is of the order of the Schwarzschild radius $r_g$. There is a finite probability of the escape of one of the particles of the pair as a real particle with positive energy. This process represents Hawking radiation. According to Hawking a black hole with mass   radiates as a black body at the temperature  $T_H = \frac{\hbar c^3}{8 \pi k_B G m}$ .  A black hole whose temperature is above ambient temperature (about $2.7 K$) should emit energy in the form of photons and other elementary particles, thereby reducing its mass and increasing temperature. An increase in temperature, in turn, leads to an increase in the speed of radiation. The black hole will inevitably come close to the Planck scale. What is her future fate? In this case, the question naturally arises: do black hole remnants  exist or does the process continue until they completely evaporate? A convincing argument against the existence of remnants is the absence of a mechanism that blocks the evaporation process when moving to Planck scale. At the same time, we must not forget that the transition to the Planck scale leads to the fundamental need for the synthesis of quantum mechanics and gravity. As we saw above at a phenomenological level, this synthesis can be realized through the transition from HUP to GUP.
\begin{figure}
 \centering
  \includegraphics[width=7 cm]{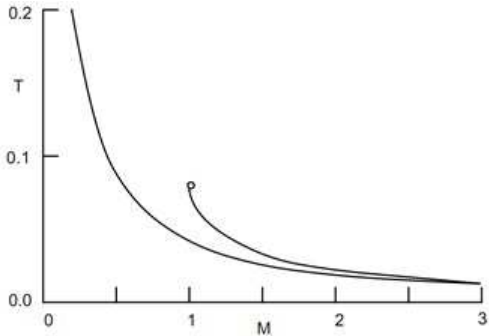}\\
  \caption{Black hole temperature as a function of its mass. Mass is given in units of Planck mass, and temperature in units of Planck temperature. The lower curve is the Hawking temperature, and the upper curve (with $\circ$) is the result using the GUP \cite{jad}.}\label{fg3}
\end{figure}

Let us turn to the useful analogy between a black hole and an atom \cite{jb,jb1}. The energy of electron in the hydrogen atom is $E = p^2 / 2 m - e^2 / r$. The classical minimum of energy at $p = r = 0$ due to the uncertainty principle cannot be realized in a quantum system. If we impose the condition $p \sim \hbar / r$, the saturation of the HUP will be achieved at $r_{min} \neq 0$.

How the transition from HUP to GUP will affect the evolution of an “atom” (a black hole) \cite{jad}?

To answer on this question we may use the GUP to derive a modified black hole temperature \cite{jad}. solving  the equation
\[\Delta x = \frac{\hbar}{\Delta p} + l_{Pl}^2 \frac{\Delta p}{\hbar}\]
we find
\[\frac{\Delta p}{\hbar} =\frac{\Delta x}{2 l_{Pl}^2} \left[1\mp \sqrt{1-\frac{m_{Pl}^2}{\Delta x^2}}  \right]\]
Assuming  $\Delta x =r_g =2 G m /c^2$ for temperature of radiated photons (Hawking temperature) we obtain \cite{jad}
\[T_{H} =\frac{\hbar c^{3} }{8\pi Gm} =\frac{m_{Pl}^{2} c^{2} }{8\pi m} \to T_{GUP} =\frac{mc^{2} }{4\pi } \left[1\mp \sqrt{1-\frac{m_{Pl}^{2} }{m^{2} } } \right]\]
Only the solution corresponding to the minus sign is physical, since it reproduces the correct value of the Hawking temperature for $m\gg m_{Pl}$.

As was natural expect the temperature to become unphysical for  $m< m_{Pl}$ or $r_g < l_{Pl}$. Derivative $dT/dM$ at $m \to m_{Pl}$ goes to infinity, which corresponds  to the zero heat capacity of the black hole.
Black hole temperature as a function of its mass is shown in Fig.\ref{fg3}.

\begin{figure}
  \centering
  \includegraphics[width=7 cm]{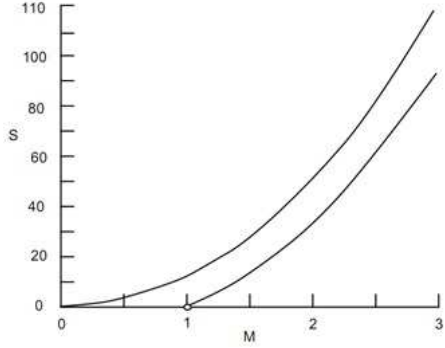}\\
  \caption{Entropy of a black hole as a function of mass (in units of Planck mass). The upper curve is the Beckenstein  entropy  $S_B$, the lower curve is the modified entropy $S_{GUP}$ \cite{jad}.}\label{fg4}
\end{figure}

The black hole entropy can be obtained by integrating the relation $dS=\frac{c^{2} }{T} dM$. If in the calculation as a temperature we use the Hawking temperature $T_{H}$, we get the standard Beckenstein  entropy $S_B$  . Temperature use $T_{GUP} $ leads to a modified entropy $S_{GUP}$,
\[{S_{B} =\frac{4\pi GM^{2} }{\hbar c} =4\pi \frac{m^{2} }{m_{Pl}^{2} } ,} \]
\[{S_{GUP} =2\pi \left[\frac{m^{2} }{m_{Pl}^{2} } \left(1-\frac{m_{Pl}^{2} }{m} +\sqrt{1-\frac{m_{Pl}^{2} }{m} } \right)-\log \left(\frac{n+\sqrt{m^{2} -} m_{Pl}^{2} }{m_{Pl} } \right)\right]}\]
The entropies $S_{B} $  and $S_{GUP}$  are shown in Fig.\ref{fg4}. Note that the modified entropy $S_{GUP}$  vanishes at $m=m_{Pl}$.

\subsection{The holographic principle}

When we talk about gravity, it’s hard not to mention the universality of this force, which acts on everything that has energy. Such universality means that its nature is connected to deep time-space properties. This pushed Einstein to create GR and to understand gravity as a result of curvature of space-time. in a certain sense, space is inseparable from gravity. We can say that GR was created on the basis of the great axiomatic theory Geometry.

There are two forces types in nature. One is connected to the presence of fields or force carriers, , another does not possesses such nature and is more likely to emergence character, appearing due to pure statistic reasons. In particular, entropic forces are one of them, connected to the tendency of of the system towards the condition with maximum entropy value. These forces are managed by another great asymptotic theory - thermodynamics.

Thermodynamics divide quantities into microscopic and macroscopic ones. It says nothing of microscopic conditions except for their equipartition. Macroscopic variables already are observable quantities. Such a deep division of quantities is especially attractive for a unified theory that would explain the appearance of space-time and matter.
A new point of view regarding gravitation  appeared lately and is based on information. It gives a completely different meaning even to the space, which can be seen as a device for keeping information about the particles conditions and their movement. A holographic principle plays a very important role in this approach. This principle is based on several sources including $AdS/CFT$-accordance \cite{hsl} and the black hole theory \cite{hs2}.

According to the field theory, the holographic principle appeared from the establishment of some unexpected correspondences between completely different theories. Let there be  manifold $M^{D+1}$ with a
boundary surface   $N^{D}=\partial M^{D+1}$. The holographic principle postulates a direct connection between a certain field theory on $M^{D+1}$ and another field theory on $N^{D}$. Wherein there is gravity in  $M^{D+1}$ while there is none gravity on  $N^{D}$. Supergravity in the anti-de Sitter space on the $M^{D+1}$ and conformal field theory on its boundary  $N^{D}$ ($AdS / CFT$ correspondence \cite{hsl}) are  the most well-known examples. There are not a lot of examples of such equivalences. In other words, the holographic principle states that the universe is similar to a hologram and the universe is perceived as three-dimensional by us can be equivalent to the quantum fields, functioning on a remote two-dimensional surface. This idea echoes the ancient ideas of Plato about illusory nature of our world in a certain way.

Below we will discuss a simpler form of the holographic principle, closely related to storing information of a 3-dimensional object on a 2-dimensional surface. Let's try to understand the appearance of this principle from the black hole physics.

If you observe a black hole from the outside, it is obvious that our ignorance of its insides is maximum. We, even in principle, cannot look beyond the horizon, and in a certain way the area beyond the horizon does not even exist in our universe. From that viewpoint the horizon separates us from an area of maximum entropy. It may be considered that inside a black hole all microscopic states become equally probable.
Microscopic states may mean everything, including space. If we would throw extra entropy in a black hole, it would have to expand and as a result, increase the horizon area. Thus it means that the amount of information consumed by a black hole is limited. With an infinite number, this is not required (see Section 2). Based on the gedanken  experiment of throwing entropy into a black hole, Beckenstein established \cite{hs2} the dependence of the black hole entropy in the following form:
\begin{equation}\label{hol3}
    S_{BH} = \frac{A}{4 l_p^2} \quad
\end{equation}
This leads us to the natural formulation of the holographic principle: the maximum entropy in volume V, which is limited by surface area $A$, is determined by relation (60).

\begin{figure}
  \centering
  \includegraphics[width=7 cm]{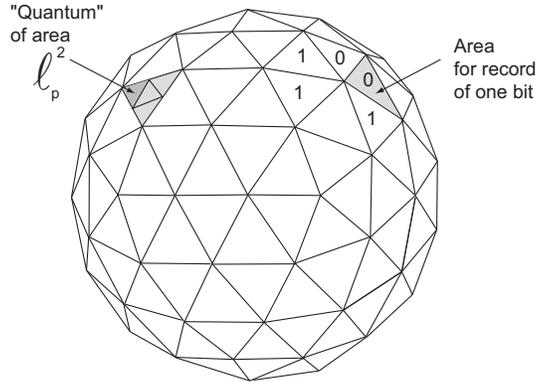}\\
  \caption{The symbolic image of the event horizon of a black hole. Planck area $l^2_p$ is shown. Area $4 l^2_p$ is required to record one bit of information.}\label{fg5}
\end{figure}

It means that the absorbed  information (we remind that information is unthinkable without physical storage) is coded on the surface of black hole's horizon. One bit of information occupies a quant of $4 l_p^2$ area of the surface of the black hole's horizon (see Fig.\ref{fg5}). In this sense, the classical theory of gravity in volume can be reformulated as a quantum theory with quantum degrees of freedom on the surface boundary. It may be said that the fundamental freedom degrees are classic, and quantumness on the surface appears as a consequence of information loss \cite{56s}. Therefore, in a finite volume, only a finite number of states can be placed and accordingly there is a limit to the amount of information that is placed in him.

An important question appears - what does the information recorded on the horizon describe? The most natural answer is everything. The fact is that space, time and matter are not the primary objects. The goal of the final theory is to explain the appearance of these objects. That is why in microscopic sense there are no space, no time and no particles. There is only a fluctuating vacuum. All these microscopic degrees of freedom are recorded in a roughened way on a holographic screen. On one side there are space. time and matter represent macroscopic concepts, on the other - microscopic states, which are mostly unknown to us. We can talk of positions and velocities of mass, that are macroscopic quantities from this point of view, before entering the horizon. It may be said that everything is possible beyond the horizon.

Thus, the amount of information proportional to the screen area can be recorded on the horizon
\begin{equation}\label{holp8}
    N=\frac{A}{l^2_p} \quad
\end{equation}
where $N$ is the number of bits that can be written on the screen. $s$ we noted, information is not conceivable without material carriers or their universal substitute - energy. Therefore, when recording information on the surface of the horizon, energy must inevitably be distributed on it. There is a natural assumption about its equipartition by bits of information. This distribution of energy, in turn, leads to the introduction of horizon temperature. Then the presence of degrees of freedom at the horizon or holographic  screen is a fairly natural property, given the temperature of the screen. According to Boltzmann's statement: If something can be heated, then it must have a microstructure. Therefore, if the holographic screen can be heated, then it required to have internal degrees of freedom.

Let us now return to the principle of equipartition among the degrees of freedom. Application of this principle to the black hole horizon is discussed in \cite{hs4,hs5}. The equipartition rule defines the relationship between energy and temperature according to
\begin{equation}\label{holp9}
    E=\frac{1}{2} N k_{B}T \quad
\end{equation}
Here  $E$  is the total energy inside the black hole or behind the holographic screen, $T$ is the temperature of the screen. According to the first law of thermodynamics, this energy is closely related to  the entropy of the screen, and therefore the properties of entropy can be transferred to  equipartition energy by bits.

After discussing the origin of the holographic principle associated with black holes, let's transfer this principle to the space-time domain even in the absence of such objects, like an event horizon. The use of the holographic principle leads to interesting conclusions not only in the field of general relativity, but also for such fundamental classical laws, like Newton’s laws \cite{58s}. Surprisingly, thermodynamics, gravity, quantum and classical theories are interwoven in the following presentation. Usually speaking about the holographic principle and its applications, use its relativistic form. Indeed, one can obtain field equations of $GR$, even for a wider class of theories, from purely thermodynamic considerations (see, for example, \cite{20s,hs8}). However, the forces of inertia and gravity are the most important attribute of our nonrelevantivistic world. Therefore, following \cite{58s}, it is convenient to begin with the nonrelativistic case. Basis holographic view of the world are ideas about holographic screens on which records information about energy or matter and ultimately about space. The question of the recording method and the physics of this process will be left open. With some limit point of view, space in the holographic world is secondary essence the primary is information and its recording on holographic screen.

Let's understand how some classical laws look from this point of view. It is natural to begin the discussion with holographic screens as the primary object. The space appears only due to the presence of its description on the holographic screen. Only macroscopic information available to an external observer is recorded on the screen. With this approach, space has only a macroscopic meaning.
\begin{figure}
  \centering
  \includegraphics[width=5 cm]{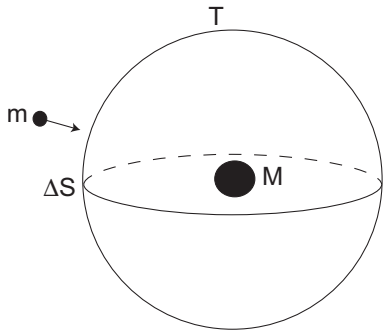}\\
  \caption{A spherical holographic screen and a particle or mass $m$ near it. Inside may be some mass $M$.}\label{fg6}
\end{figure}

Thus, on one side of the holographic screen there is already macroscopic (phenomenological) space. There is no space on the other side of the screen. There are only microscopic states necessary for its creation. Strictly speaking, microscopic details are not accessible to the observer.

In order to understand the connection between dynamics and information describing it, we consider a particle of mass $m$ that approaches a holographic screen (see Fig.\ref{fg6}). Information on this particle should in some way be reflected in holographic screen. Remembering the properties of the event horizon, holographic  screen should be attributed to such a fundamental property as temperature. Postulating that the screen is in thermodynamic equilibrium, it should be assumed that the temperature of the holographic screen everywhere the same. The entropy of the screen is finite, as is the information recorded on him. Therefore, for simplicity and convenience, we will represent the holographic screen as a closed surface. The final information is recorded on such a surface about the inner space. Consider now another closed holographic surface embedded in the first. The information recorded on it is less than the initial one by the amount of information about the layer of space between them. A peculiar holographic foliation arises related to entropy

Let us now return to the recording of information about the mass $m$  near the holographic screen. Using Beckenstein’s conception for introducing the entropy of a black hole, it can be postulated that the change in screen entropy associated with the displacement of a particle $\Delta x$  is determined by the expression \cite{58s}
\begin{equation}\label{holp5}
    \Delta S =2 \pi k_{B} \frac{mc}{\hbar} \Delta x \quad
\end{equation}
This determination can be considered as a linear approximation by the particle displacement. Proportionality to mass $m$  provides additive of entropy. Constants $\hbar$, $c$ provide correct dimension.  An approach of the particle to the screen should be slow enough to maintain thermodynamic equilibrium on the screen.

We introduce now the important concept of entropy force. Using the thermodynamic definition of entropy we introduce the force $F$,
\[F \Delta x = T \Delta S \quad  \]
If we use the postulate of a change in entropy (\ref{holp5}), then an entropic force arises, the value of which is determined by the temperature of the screen
\begin{equation}\label{holp6}
    F  = T \cdot 2 \pi k_{B} \frac{mc}{\hbar} \quad
\end{equation}
Thus, this entropic force is nonzero at a nonzero temperature $T$. Now recall that in the uniformly accelerated reference frame the Unruh temperature is observed (see \cite{hs10})
\[k_{B} T=\frac{\hbar a}{2 \pi c} \quad .\]
Here $a$ is a acceleration. Substituting this $T$ Unruh temperature in relation (\ref{holp6}), we obtain Newton's second law
\begin{equation}\label{holp7}
    F=m a \quad
\end{equation}
In this case, Unruh temperature is understood as the temperature of the screen at which e mass $m$ reaches  acceleration equal to  $a$.

Let's move  to the discussion of gravity from a holographic point of view. Let a mass  $M$  be placed inside a spherical screen (see Fig.\ref{fg6}). Therefore, energy $E=M c^2$  is concentrated in this region. Using relations(\ref{holp8}) and (\ref{holp9}), we obtain for temperature
\[k_B T= \frac{2 M c^2 l^2_{pl}}{A} \quad  \]
Substituting this temperature into relation (\ref{holp6}) for the entropic force and considering that screen area  $A =4 \pi r^2$ we get the great law of gravity
\begin{equation}\label{holp11}
    F= G \frac{m M}{r^2} \quad
\end{equation}
Of course, the most important thing in this way of deriving this law is changing the interpretation of the gravitational force. From this point view of gravitational force loses the status of fundamental force and takes on meaning entropic, statistical force \cite{58s}. Such a view of gravitational force is well compatible with observable facts.  The main property of gravity is its universality. Gravity acts equally on everything that has energy, regardless of the scale of distances between different material objects.

Of course, the non-relativistic physics discussed above should be transformed into   relativistic. One of the first attempts to explain GR from thermodynamic considerations can be found in \cite{20s}. In this work, Einstein's equations were are derived from the fundamental relation $dQ =T dS$ and proportionality of entropy to horizon area. Appendix  demonstrates the derivation of Einstein's equations in the framework of thermodynamic formalism.

\subsection{Discrete space structure and holography}

The traditional point of view suggested that the dominant part of the degrees of freedom of our world are fields filling space. However, it gradually became  clear that such a representation complicates the construction of quantum field theory. To give meaning to the latter, cutting off all integrals included in the theory at small distances is required.  Formally, this difficulty can be avoided by going to the description of our world on a discrete spatial lattice. The period of such a lattice remains a free parameter of the theory, which should be determined in future microscopic theories.

Recently, some physicists have taken an even more radical point of view: instead of a three-dimensional lattice, a full description of nature requires only a two-dimensional lattice located on the spatial boundary of the studied region \cite{56s,57s}. Such the approach is based on the so-called “holographic principle”. The name of the principle is associated with an optical hologram representing a two-dimensional record of a three-dimensional object. The holographic principle is based on the relationship between entropy and information.

The amount of information $I$ associated with matter and its position is measured in terms of the entropy $S$,
\begin{equation}\label{e3_24}
  \Delta S =  - \Delta I
\end{equation}
A change of entropy with the displacement of matter leads to the entropic force [see section 4.2]. Therefore, its origin is based on the universal tendency of any microscopic system maximize entropy. Fundamental fields associated with entropic forces are absent and dynamic equations are directly expressed in terms of changes of entropy.

As we saw above, according to the holographic principle, the theory on the spatial boundary of the volume under study should contain no more than one degree of freedom per Planck area or, in other words, the total number of degrees of freedom N obeys the inequality
\begin{equation}\label{e3_25}
  N \le \frac{A}{{l_{Pl}^2}} = \frac{{A{c^3}}}{{G\hbar }}
\end{equation}
(we omit numerical factors of the order of unity) This means that the density of information on a holographic screen is limited to ${10^{69}}bit/{m^2}$. The holographic principle as a strict statement is valid only for black holes. In other cases, it is just a hypothesis. Therefore, the answer to the question is of interest: is the holographic principle valid in a theory that combines quantum mechanics and gravity? On the one hand, such a theory generates minimum length and provides a natural transition to a three-dimensional spatial lattice. But, on the other hand, for the validity of the holographic principle, it is required that all the bulk information of such a three-dimensional lattice can be recorded in a two-dimensional lattice. In other words, such a lattice must be at least effectively two-dimensional.

Let's make sure that the inclusion of gravitational effects provides effective two-dimensionality \cite{59s}. To do this, consider some three-dimensional volume with a characteristic linear size $l$.  In theory with a minimum length ${l_{\min }} \simeq {l_{Pl}}$  it is natural to assume that the volume is divided into cells of magnitude  ${\left( {{l_{Pl}}} \right)^3}$ . However how shown above (see section 2.3), this conclusion is erroneous. By virtue of a fundamental limitation, values  $\delta l \ge {\left( {ll_{Pl}^2} \right)^{1/3}}$   the minimum size of the cell into which we can split the cube with the edge  is  ${\left( {ll_{Pl}^2} \right)^{1/3}}$   (see Fig.\ref{fg6}). In other words, space-time foam leads to the formation of a dynamic lattice with
\begin{equation}\label{e3_26}
  N = {\left[ {\frac{l}{{{{\left( {ll_{Pl}^2} \right)}^{1/3}}}}} \right]^3} = \frac{{{l^2}}}{{l_{Pl}^2}}
\end{equation}
If we associate with each cell one degree of freedom of the system, then the resulting ratio reproduces the holographic principle in the form of (\ref{e3_25}).

Note that although the original formulation of the holographic principle was made in terms of information, its form (\ref{e3_26}) is purely dynamic. A similar situation is typical for physics as a whole.

Thus, there are two ways to describe reality: in terms of dynamics (dynamic variables) and in terms of statistical physics, where the key role is played by entropy directly related to information. Description methods do not exclude but complement each other.
\begin{figure}
  \centering
  \includegraphics[width=5 cm]{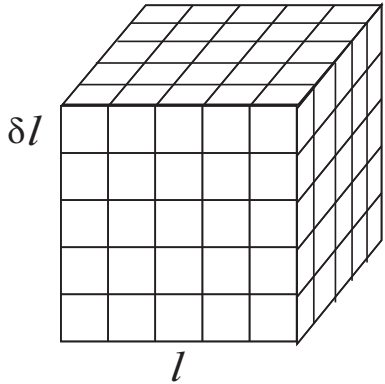}\\
  \caption{The division of volume into unit cells.}\label{fg7}
\end{figure}

If we have two ways of description of reality, then the question always arises of the “equivalence” of principles that lie at their core. In other words, there must be a translation dictionary from one language to other. This means that there is some correspondence between the terms of one and the other approach. We use the term "equivalence" to understand the following statement: all the results of one approach can be obtained in another. Of course it’s very  long way. And there is always the possibility that some new observation will turn out to be a stumbling block for one of the approaches, and then we will find the “main approach". An alternative approach is direct proof of the identity of the principles, which consists in obtaining 1 of 2 and vice versa.

Proof of equivalence of fundamental principles or attempt to find the “main” principle is dictated by the desire to insure yourself in case between principles will cause contradictions. But such contradictions have already happened and led to crises. It is enough to recall the problem with a long history: reversibility mechanics and the arrow of time in thermodynamics.

In order to avoid potential contradictions, let us consider under such in terms of the principle of maximum force, based on a dynamic approach, and the holographic principle of informational (entropic) nature \cite{60s}.

A priori, the principle of maximum force and the holographic principle are related by the fact that each of these principles is a statement about the existence of some limit value: limit force (or power) in the first case, and limit information density in the second case. In addition, a comparative analysis of these of principles draws attention to the common source of their origin - the event horizon. Of additional interest is the fact that principles may be implemented at significantly different scales, and the proof of their equivalence can be seen as further confirmation of IR-UV correspondence \cite{coh}.

First we go through the intended path in one direction, we obtain the principle of maximum force from the holographic principle, or, in other words, try to obtain the value of the maximum power   using the fundamental holographic limitations.

As we saw above, the total number of elementary logical operations that
the system can execute per unit time, limited by the average excess of energy above ground state  ${N_{ops/\sec }} \le \frac{{2E}}{{\pi \hbar }}$, and the total number of bits with which the system can work is limited by its entropy  ${N_{bits}} \le S/{k_B}\ln 2$.

To transform information into energy, we use the Landauer principle \cite{39s}, according to which in any computer system, regardless of its physical realization, for creating one bit of information energy is needed
\begin{equation}\label{e3_27}
  {E_{bit}} \ge {E_{SNL}} = {k_B}T\ln 2
\end{equation}
Here ${E_{SNL}}$ is Shannon-von Neumann-Landauer energy. For $T = 300 \,K^{\circ}$ energy  ${E_{SNL}}\, \approx 0.017 \, eV$. Relation (\ref{e3_27}) is equivalent to the statement that the temperature can be considered the average energy of one bit of information on a holographic screen. Considering the creation of one bit of information as an elementary logical operation, we can write for the maximum power $N_{max}$ spent with the energy reserve $E$   and the temperature of the holographic screen $T$,
\begin{equation}\label{e3_28}
  {N_{\max }} \approx {N_{ops/\sec }} \times {E_{SNL}} \approx \frac{E}{\hbar }{k_B}T
\end{equation}
As we saw above, limit values are reached only on the event horizon. Using the Hawking temperature ${T_H} = \frac{{\hbar {c^3}}}{{8\pi {k_B}GM}}$  as the horizon temperature and $E = M{c^2}$ we get
\begin{equation}\label{e3_29}
  \eta  \approx \frac{{{c^5}}}{G}
\end{equation}
which exactly corresponds to the value of the maximum power. Now do the reverse path i.e. we show that the maximum power $\eta  \approx \frac{{{c^5}}}{G} \approx {10^{52}}W$
does not allow to achieve information density exceeding  $l_{Pl}^{ - 2} \approx {10^{69}}bit/{m^2}$.

As shown in \cite{21s}, a direct consequence of the principle of maximum force is so-called the horizon equation
\begin{equation}\label{e3_30}
  \delta E = \frac{{{c^2}}}{{8\pi G}}a\delta A
\end{equation}
where $a$ is surface gravity associated with the horizon temperature $T$ by the relation
\begin{equation}\label{e3_31}
  a = \frac{{2\pi {k_B}}}{{\hbar c}}T
\end{equation}
The energy flux $\delta E$, the intersecting element of the horizon area  $\delta A$ obeys the balance equation \cite{20s}
\begin{equation}\label{e3_32}
  \delta S = \frac{{\delta E}}{T}
\end{equation}
Using relations (\ref{e3_30}) and (\ref{e3_31}) we obtain the holographic principle in differentia form
\begin{equation}\label{e3_33}
  \delta S = \frac{{{c^3}}}{{4G\hbar }}\delta A = {k_B}\frac{{\delta A}}{{4l_{Pl}^2}}
\end{equation}
Hence, after substituting ${N_{bits}} \le S/{k_B}\ln 2$, we find the restriction  ${N_{bits}}/A \le l_{Pl}^{ - 2} \approx {10^{69}}bit/{m^2}$  in full accordance with the limit recording density on the holographic screen.

\subsection{Acceleration limit}

The condition for the existence of traditional space-time in the presence of vacuum polarization processes (virtual processes of pair production and annihilation caused by quantum fluctuations) leads to the limitation of proper acceleration relative to vacuum or, in other words, to the appearance of maximum acceleration \cite{63s,64s,65s}.

The proper acceleration of the particle  $a$  in curved space-time is scalar quantity, which is determined by the ration
\begin{equation}\label{3_34}
  {a^2} =  - {c^4}{g_{\mu \nu }}\frac{{D{v^\mu }}}{{ds}}\frac{{D{v^\nu }}}{{ds}}
\end{equation}
where ${g_{\mu \nu }}$ is the metric tensor, ${v^\mu } \equiv d{x^\mu }/ds$  is the dimensionless 4-velocity of the particle, and  $D/ds$ is the covariant derivative with respect to the linear element of the particle’s world line,
\begin{equation}\label{e3_35}
  \frac{{D{v^\mu }}}{{ds}} \equiv \frac{{d{v^\mu }}}{{ds}} + \Gamma _{\alpha \beta }^\mu {v^\alpha }{v^\beta }
\end{equation}
Here $\Gamma _{\alpha \beta }^\mu$ is Christoffel symbol. Under the action of only the gravitational field, the particle moves along the geodesic with zero proper acceleration. At the inclusion of non-gravitational forces, the acceleration is different from zero. Using simple physical arguments, we show that the maximal value of this acceleration is a fundamental property of space-time.

We give these arguments following \cite{66s}. From the uncertainty principle energy-time it follows that the lifetime of the virtual particle-antiparticle pair (with particle mass $m$) , arising due to vacuum fluctuations $ \approx \hbar /2m{c^2}$, and the distance traveled during this time is $ \approx \hbar /2mc$  (Compton pair wavelength). If a particle gains energy equal to its rest mass, the virtual particle transforms into a real one. Turning in particle rest system, which is generally non-inertial, we will find that the inertial force acts on the particle is ${F_{in}} = \left| {ma} \right|$, where $a$ is the acceleration of the particle. The work  $A$  performed by the inertial force of during the particle’s lifetime is equal to  $A = ma \times \frac{\hbar }{{2mc}}$. If $A = m{c^2}$ then
\begin{equation}\label{e3_36}
  a = \frac{{2m{c^3}}}{\hbar }
\end{equation}
With this acceleration, particles of mass $m$ will be produced in abundance from vacuum. Further increase in acceleration will lead to an increase in the mass of the born particles. What critical consequences can an unlimited increase in acceleration lead to? For big enough accelerated particles can transform into black holes. This will happen when the Compton particle length $\hbar /mc$ becomes less than its gravitational radius,
\begin{equation}\label{e3_37}
  \hbar /mc < \frac{{2Gm}}{{{c^2}}}
\end{equation}
It follows that the threshold for the formation of black holes is the mass of the order of Planck mass ${\left( {\hbar c/G} \right)^{1/2}}$. Substituting $m = {m_{Pl}}$ into relation (\ref{e3_36}), we find
\begin{equation}\label{e3_38}
  {a_0} \approx {\left( {\frac{{{c^7}}}{{\hbar G}}} \right)^{1/2}}
\end{equation}
(We still omit factors of the order of unity) With this acceleration, birth black holes with Planck mass due to vacuum polarization will lead to destruction  the traditional idea of the structure of space-time, and the concept of acceleration itself will lose its usual meaning.

For this reason, $a_0$   should be considered as maximum proper particle acceleration relative to vacuum. Consider now the mechanism of the appearance of maximum acceleration at the microscopic level. As shown
Schwinger \cite{67s} probability $P$ of production from vacuum under the influence of an external electric field $E$ of an electron-positron pair per unit time in unit volume is
\begin{equation}\label{e3_39}
  \frac{{{d^4}P}}{{dtdV}} = \left( {\frac{{{e^2}{E^2}}}{{{\pi ^2}{\hbar ^2}c}}} \right)\sum\limits_{n = 1}^\infty  {\frac{{{e^{ - n\frac{{{E_c}}}{E}}}}}{{{n^2}}}}
\end{equation}
Here, $m$ is the electron mass and $E_c$ is the characteristic electric field representing the threshold effective pair production,
\begin{equation}\label{e3_40}
  {E_c} = \pi {m^2}\frac{{{c^3}}}{{e\hbar }}
\end{equation}
How to estimate the critical field $E_c$? Using the energy-time uncertainty principle, we find that a virtual electron-positron pair lives in a vacuum
\begin{equation}\label{e3_41}
  \Delta t \sim \frac{\hbar }{{\Delta E}} \sim \frac{\hbar }{{2m{c^2}}}
\end{equation}
Fluctuation spreads over a distance
\begin{equation}\label{e3_42}
  \Delta x \sim c\Delta t \sim \hbar /2mc
\end{equation}
If over time $\Delta t$  an external field $E$ acting with a force $e E$ on each of the virtual particles will do (at a distance $\Delta x$ ) work
\begin{equation}\label{e3_43}
  \left( {eE} \right)\left( {\hbar /2mc} \right) \sim m{c^2}
\end{equation}
the virtual particle is transformed into the real one. From relation (\ref{e3_43}) we find
\begin{equation}\label{e3_44}
  {E_c} = \frac{{2{m^2}{c^3}}}{{e\hbar }} \sim \frac{{\pi {m^2}{c^3}}}{{e\hbar }}
\end{equation}
Of course, if instead of electric force  we use the inertial force we
reproduce the result (\ref{e3_38}) It is well known \cite{69s} that an observer in a uniformly accelerated system with acceleration $a$ will record the radiation generated by vacuum with temperature $T$ (Unruh temperature)
\begin{equation}\label{e3_45}
  T = \frac{{\hbar a}}{{2\pi {k_B}c}}
\end{equation}
Substituting relation (\ref{e3_38}) into (\ref{e3_45}) we find the maximum possible temperature Unruh (Sakharov temperature \cite{71s,72s})
\begin{equation}\label{e3_46}
  {T_U}_{\max } = \frac{{{{\left( {\frac{{{c^5}}}{G}\hbar } \right)}^{1/2}}}}{{{k_B}}}
\end{equation}
Therefore, the observer in the system with maximum acceleration ($\left( {{a_0} \approx {l_{pl}}/t_{pl}^2 \approx {{10}^{52}}m/{{\sec }^2}} \right)$) will be surrounded by thermal radiation with a temperature  ${T_U}_{\max } \approx {T_{Pl}} \approx {10^{32}}K$. The result (\ref{e3_46}) can be given a more general form,  remembering the definition of maximum power $\eta  = {c^5}/G$,
\begin{equation}\label{e3_47}
  {k_B}{T_U}_{\max } = {\left( {\frac{{{c^5}}}{G}\hbar } \right)^{1/2}} = {\left( {\eta \hbar } \right)^{1/2}}
\end{equation}
We see that the existence of maximum power predetermines the Sakharov temperature. Both the first and second restrictions have a common source - the event horizon.

Note that the first to draw attention to the problem of maximum acceleration E.Caianiello \cite{73s}. The value of the maximum acceleration obtained by him ${a_{\max }} = {c^2}/\lambda $, where $\lambda $ is linear particle size. Substituting $\lambda  = {l_{Pl}} = {\left( {\hbar G/{c^3}} \right)^{1/2}}$ we will receive immediately (\ref{e3_38}). The result attracted attention, since the presence of finite maximum acceleration eliminated a number of undesirable infinities. However, the method of obtaining it caused some doubts. Somewhat later \cite{74s} Caianiello reproduced the initial result using the energy-time uncertainty principle in the form
\begin{equation}\label{e3_48}
  \Delta E\Delta f(t) \ge \frac{\hbar }{2}\left| {\frac{{df}}{{dt}}} \right|
\end{equation}
where $f(t)$ is an arbitrary differentiable function. Since we are interested in the maximum acceleration, it is natural to consider the initially resting particle and choose $f(t) = v(t)$.  Then, provided that $\Delta E \le E,\;\Delta v \le v \le c$, we obtain
\begin{equation}\label{e3_49}
  \frac{\hbar }{2}\left| a \right| \le \Delta E\Delta v \le m{c^2} \times c
\end{equation}
From here we immediately will find (\ref{e3_36}) and when substituting $m = {m_{Pl}}$, we reproduce the fundamental result (\ref{e3_38}).

The presence of maximum acceleration $a_0$ automatically leads to the  existence of a minimum radius of curvature ${R_{\min }}$ along the world particle lines.  The radius of curvature of the world line $R = {c^2}/a$. Therefore, the minimum radius of curvature
\begin{equation}\label{e3_50}
  {R_{\min }} = \frac{{{c^2}}}{{{a_0}}} \approx {\left( {\frac{{\hbar G}}{{{c^3}}}} \right)^{1/2}} = c{\left( {\frac{\hbar }{\eta }} \right)^{1/2}}
\end{equation}
Once again, we are convinced of the key role of the horizon, which generates maximum power and, as a result, the maximum proper acceleration and the minimum radius of curvature of the world line.

\section{Cosmological aspects of the physics of limit values}

In all man-made experiments performed with everything currently available accuracy, space-time appears continuous and structureless. Such a result  should not be surprising. The achieved energies $\sim 10$ TeV and spatially-temporary permissions ($\sim 10^{-16}$ cm, $\sim 10^{-16}$ sec)  are too far from the Planck scales.  In such  situation, in search of quantum manifestations of classical gravity, it is natural to turn to a unique object --- the Universe, which, during its evolution, has passed a gigantic interval of energies and spatial scales.

\subsection{Relationship of large and small scales (IR-UV compliance)}

Attempts to discover the connections between the extremely large and small numerical characteristics of the world around us are an exciting game of chance. Accidental coincidences provoked intense enthusiasm and the illusion of unraveling of the hidden secrets nature.  The roots of the game go back to antiquity. First "Theory of Numbers" is Kabala which denied the possibility of random coincidences, considering the numbers a symbol of spiritual nature of things. t is interesting to note that already in 1923, in an appendix to his book "Space, Time, Matter", German Weil estimated the ratio of the radius of the World to the radius of an electron like $\sim 10^{40}$.
As he writes, this suggests that the enormous numerical value of the constant $c^{4} /8\pi G$  is related to the difference in the sizes of the electron and the Universe. P. Dirac belongs observation that in a hydrogen atom the ratio of electric forces to gravitational is close to the ratio of the size of the Universe to the size of the electron.

Nevertheless, until recently (at a phenomenological level), large and small scales were considered independent. The situation is changing radically if we take into account that even macro objects have quantum properties. This connection facilitates the solution of some fundamental problems in particular, the problems of the cosmological constant. Explain the essence of this problem. To describe the observed dynamics of the universe, we are forced suggest that the dark energy providing the observed accelerated expansion of the universe represents the dominant component. Therefore, we can, using the first Friedman equation, estimate its density,
\begin{equation} \label{ek1_1}
  \rho _{\Lambda } \approx \frac{H_{0}^{2} m_{Pl}^{2} }{8\pi } \approx 10^{-47} GeV^{4}
\end{equation}
According to the most popular version, dark energy in the form of a cosmological constant you $\Lambda $ represents the zero-point oscillations of the quantized fields and, therefore, within such an interpretation
\begin{equation} \label{ek1_2}
\rho _{vac} =\frac{1}{2} \int _{0}^{\infty }\frac{d^{3} k}{\left(2\pi \right)^{3} } \sqrt{k^{2} +m^{2} }  =\frac{1}{4\pi ^{2} } \int _{0}^{\infty }dkk^{2} \sqrt{k^{2} +m^{2} }
\end{equation}
The integral diverges (ultraviolet divergence) because $\rho _{vac} \sim k^{4} $. Assuming that in quantum field theory there is a certain cutoff scale $k_{max}$ , we make the integral (\ref{ek1_2}) finite
\begin{equation} \label{ek1_3}
  \rho _{vac} \approx \frac{k_{\max }^{4} }{16\pi ^{2} }
\end{equation}
The natural choice seems $k_{\max } = m_{Pl} =1.22\times 10^{19} GeV \,(\hbar = c =1)$, since it is customary to identify the boundary of applicability of GR with this quantity. The result of this choice will be the density of vacuum energy
\begin{equation} \label{ek1_4}
   \rho _{vac} \approx 10^{74} \, GeV^{4}
\end{equation}
The obtained value of the energy density exceeds the observed value by more than 120 orders of magnitude (!).This contradiction presents the so-called problem of the cosmological constant \cite{pa0,pa1,pa2}. Physics has never encountered such a giant numerical contradiction.

Before the discovery of the accelerated expansion of the Universe, efforts were mainly focused on the search for universal symmetry, leading to the strict zeroing of the cosmological constant. Owever, despite decades of research, no mechanism has been found to ensure consistent implementation of this requirement. It is surprising how it can to slip away such a powerful mechanism, reducing a gigantic number of significant digits? After the discovery of accelerated expansion, the attitude towards the cosmological constant sharply has changed. Being the simplest mechanism, generating “anti-gravity” she attracted everyone's attention. Now the main efforts were not directed at her “zeroing”, but to search for reasons that make it much less than the value expected for dimensional reasons.

The discovery of supersymmetry led to the hope that, since bosons and fermions (with identical masses in the limit of exact supersymmetry) give identical contributions to vacuum means, but with opposite signs, the problem of the cosmological constant can be solved with a reasonable balance of fermions and bosons in nature. However, supersymmetry (if it exists) is obviously broken at low temperatures, ruling today in the Universe. For this reason, it can be expected that the cosmological constant was equal to zero in the early Universe, but has revived recently. However, this is an undesirable scenario, almost the opposite of the one we are looking for, since large  $\Lambda$  values in the early Universe are attractive in terms of inflation, while how very small current $\Lambda$  values are consistent with observations.

A “game” with fundamental constants allows a transition to a new set of Planck variables. Since the cosmological constant plays such a fundamental role in the dynamics of the Universe, let us consider the transition from the original  set of Planck units $m_{Pl}$, $l_{Pl}$, $t_{Pl}$, built on the fundamental constants $\hbar$, $c$, $G$  to the new set $m_{\Lambda}$, $l_{\Lambda}$, $t_{\Lambda}$, constructed on the constants $\Lambda$, $c$, $G$. New Planck units \cite{chav}
\begin{equation} \label{ek1_6}
\begin{array}{l} {m_{\Lambda } =\left(\frac{\hbar _{\Lambda } c}{G} \right)^{1/2} =\frac{c^{3} }{G\left(8\pi \Lambda \right)^{1/2} } =5.90\times 10^{56} GeV;} \\ {l_{\Lambda } =\left(\frac{G\hbar _{\Lambda } }{c^{3} } \right)^{1/2} =\left(\frac{8\pi c^{2} }{\Lambda } \right)^{1/2} =4.38\times 10^{28} cm;} \\ {t_{\Lambda } =\left(\frac{G\hbar _{\Lambda } }{c^{5} } \right)^{1/2} =\left(\frac{8\pi }{\Lambda } \right)^{1/2} =1.46\times 10^{18} s} \end{array}
\end{equation}
The new “Planck” units \cite{chav} perfectly reproduce the mass, size and lifetime of the observed Universe, but deciding the problem of the cosmological constant, give rise to the problem of Planck constant,
\begin{equation} \label{ek1_7}
  \hbar _{\Lambda } =7.35\times 10^{122} \hbar =7.75\times 10^{88} J\cdot s
\end{equation}
All the same 120 orders! Just difficulties moved to another place Apparently, fundamentally new approaches are needed to solve the problem of the cosmological constant.

It is well known that in the world there is nothing more dangerous than trying to overcome the abyss in two jumps. The concept of limit values gives us a chance, at least in  principle, by one jump to overcome the abyss of 120 orders of magnitude. In order to prepare for this jump, we briefly outline the hypothesis \cite{coh}, called $IR-UV$ (ultraviolet - infrared) correspondence.  The hypothesis is based on the following arguments

In any effective quantum field theory defined in a spatial region with a characteristic size $L$  and using ultraviolet cutoff $\Lambda$, entropy  $S\propto \Lambda ^{3} L^{3} $. For example, fermions located in the nodes of the spatial lattice of  size $L$  and  period $\Lambda^{-1}$  are in one of  states. Consequently, entropy of such a system is  $2^{(L\Lambda )^{3}}$. According to the holographic principle (see section 4) the cutoff value $\Lambda$  must satisfy the inequality
\begin{equation} \label{ek1_8}
   L^{3} \Lambda ^{3} \le S_{BH} =\frac{A}{4l_{Pl}^{2} } =\pi L^{2} m_{Pl}^{2}
\end{equation}
Here $S_{BH}$ is the entropy of a black hole with a gravitational radius $L$. We obtained an important result \cite{51s} in the framework of holographic dynamics, the value of IR-cutoff $L$ is tightly connected with the value of UV-cutoff   $\Lambda$. In other words, physics at UV-scales depends on physics parameters at IR-scale. In particular, in the case of saturation of the  inequality
\begin{equation} \label{ek1_9}
 L \sim \Lambda ^{-3} m_{Pl}^{2}
\end{equation}
From the point of view of the physics of limit values, the connection of small and large scales can be obtained from a fairly natural condition: the total energy enclosed in volume with  linear size $L$, must not exceed the mass of a black hole of the same size, i.e.
\begin{equation} \label{ek1_10}
  L^{3} \rho _{\Lambda } \le M_{BH} \sim Lm_{Pl}^{2}
\end{equation}
Here $\rho _{\Lambda}$ is the energy density in the volume $L^{3}$. In case of violation of this inequality, black hole is formed with an event horizon that prevents  further increase in energy density.

Let's consider once more possibility of finding a connection between large and small scales. The relation $\delta l=l_{Pl}^{2/3} l^{1/3}$ can be considered as the relation between the infrared and ultraviolet scales in effective quantum field theory, in which the thermodynamic laws of black holes are satisfied. In particular, this means that the entropy  $S$ of any object of linear size $l$ in such a theory should be less than the entropy of a black hole $S$ of the same size,
\begin{equation} \label{ek1_11}
  S\le S_{BH} \approx \left(\frac{l}{l_{Pl} } \right)^{2}
\end{equation}
Now consider a box of size $l$ (IR-scale) filled with a substance with UV-cutoff scale $\Lambda$. Then the entropy of such a system is $S\sim l^{3} \Lambda ^{3}$ and according to (\ref{ek1_11})
\begin{equation} \label{ek1_12}
  l^{3} \Lambda ^{3} \le \left(\frac{l}{l_{Pl} } \right)^{2}
\end{equation}
It is natural to identify the inverse UV- scale with minimal uncertainty due to length measurements, $\delta l=\Lambda ^{-1}$. In this case, relation (\ref{ek1_12}) immediately transforms in $\delta l\le l_{Pl}^{2/3} l^{1/3}$.

Returning to the problem of the cosmological constant, we apply inequality (\ref{ek1_10}) to the Universe as a whole. In this case, it is natural to identify the IR-scale with the Hubble radius $H^{-1}$, and by $\rho_{\Lambda}$ we mean the density of the dominant component filling  the Universe, i.e. dark energy in the form of a cosmological constant. Then for the top border of the energy density we find
\begin{equation} \label{ek1_13}
   \rho _{\Lambda } \sim L^2 m_{Pl}^{2} \sim H^2 m_{Pl}^{2}
\end{equation}
Given that
\[{m_{Pl} \simeq 1.2\times 10^{19} \, \, GeV;}\]
\begin{equation} \label{ek1_15}
  {H_{0} \simeq 1.6\times 10^{-42} \, \, GeV}
\end{equation}
finally get
\begin{equation} \label{ek1_16}
  \rho _{\Lambda } \sim 10^{-46} \, \, GeV^{4}
\end{equation}
This quantity is close  to the observed density of dark energy (\ref{ek1_1}). The result seems extremely interesting, but its value should not be exaggerated: it does not represent a solution to the problem of the cosmological constant, but only an indication of the direction in which this solution should be sought.

\subsection{Space-time foam and the parameters of the Universe}

Two fundamental consequences of the synthesis of quantum mechanics and gravity are the minimum length and space-time foam. The existence of the first leads to a change (on a Plank scale) of the evolution of any dynamical system, and the second gives rise to a number of universal restrictions that apply at any scale, up to the scale of the Universe.

We begin by considering the scenario for the birth of the Universe, known as "A Universe from Nothing"\cite{86s,87s}. Despite the aesthetic beauty of the name, this physical wording is fundamentally incorrect. The birth of the Universe is regarded as a quantum process. The universe is born out of vacuum exactly how a virtual particle is born. And the first and second process is governed by energy-time uncertainty principle  $\Delta E \Delta t \geq \hbar /2$.

Interpretation of this relation  requires caution, since there is no operator representing the time: We mean by $\Delta E$ the uncertainty of the energy of a micro-object in some process of duration $\Delta t$. The universe born as a quantum fluctuation in further evolving according to the laws of quantum mechanics and GR. Adequacy such  model is predetermined by answers to two simple questions. What is the lifetime of the fluctuation? Is it possible to use the fluctuation parameters as the initial conditions for SCM?

The first question is easy to answer. Quantum fluctuation lifetime comparable with the lifetime of the Universe $T_U$  can be achieved under the condition  $\Delta E \approx \hbar /\Delta t \approx \hbar /T_U$. The smallness of $\Delta E$  is ensured by compensation by the positive rest energy of fluctuations and negative gravitational energy.

To answer the second question, let us trace the temporal evolution of fluctuations. Assuming this evolution to be spherically symmetric (isotropic Universe), we represent the fluctuation energy in the form
флуктуации в виде
\begin{equation} \label{ek1_33}
   \Delta E\sim \frac{4}{3} \pi R^{3} \rho
\end{equation}
Here $\rho$ is the energy density of the fluctuation   and its radius. Let us evaluate the radius of the “Hubble bubble” as $R \sim c H^{-1}$, and the lifetime of the emerging Universe as $\Delta t\sim H^{-1}$. Using these estimates, we obtain
\begin{equation} \label{ek1_34)}
   \Delta E\Delta t\simeq \hbar /2\Rightarrow \frac{4\pi }{3} \rho \left(\frac{c}{H} \right)^{3} H^{-1} \simeq \frac{\hbar }{2}
\end{equation}
From here for mass density we find
\begin{equation} \label{ek1_35)}
   \frac{\rho }{c^{2} } \simeq \frac{3c^{5} }{8\pi G^{2} \hbar } =\frac{3}{8 \pi} \rho_{Pl}
\end{equation}
The Universe born in the framework of this scenario has a density close to Planck, and its further evolution can be considered in the framework of SCM.

The connection between physics of large and small scales can be investigated in the framework of the holographic principle, which allows us to move from space-time foam to cosmological scales. In relation to cosmology, holographic principle predicts that cosmic energy has a critical density: to avoid the formation of black holes, the energy density of the Universe must satisfy the inequality $\rho \le \left(ll_{Pl}^{2} \right)^{-1}$, where  is the linear size of  the Universe.

So far, we have considered only static space-time foam. A generalization to the case of an expanding Universe can be realized by replacing $l$ with the Hubble radius $H^{-1} $ in all "static" inequalities.

The holographic model of the Universe is based on two statements:

\begin{enumerate}
  \item Critical energy density $\rho \sim \left(H/l_{Pl} \right)^{2}$.
  \item The Universe of size $R_{H} =H^{-1} $  contains no more $HR_{H}^{3} /l_{Pl}^{2} =\left(l/l_{Pl} \right)^{2}$ bits of information.
\end{enumerate}

Using the holographic model of the Universe, we can evaluate its capabilities store and process information \cite{87bs}. All physical systems record and process make information in the course of its evolution. The laws of physics determine volume information that a physical system (system memory in bits) can record and its processing speed (number of operations per unit time). The Universe is physical system and its ultimate informational characteristics are extremely important to understand the speed of its evolution.

The critical density of the Universe in the current era according to the prediction holographic model $\left(R_{H} l_{Pl} \right)^{-2} \sim 10^{-9} J/m^{3}$  agrees well with observed average energy density.

Recall (see Section 4) that the information processing speed $\nu$ is limited by the Margolis-Levitin theorem, according to which  $\nu \le 2E/\pi \hbar$, where $E$ the available energy. For. $\rho \sim \left(H/l_{Pl} \right)^{2}$  we find for the upper limit of computation speed $\nu \sim \rho R_{H}^{3} \sim R_{H} l_{Pl}^{-2} \sim 10^{106} $. The total number of operations performed during the existence of the Universe (about 10 billion years) $\sim 10^{122}$. (We omit factors of the order of unity)

The physical meaning of this result is quite simple. Let us make a simplifying assumption: let the evolution of the Universe fully conduct during the period of dominance of matter. Due to the conservation of the number of particles in the comoving volume (and, as a result, the available energy) the speed of information processing does not change during evolution and is equal
\begin{equation} \label{ek1_36)}
   \nu \approx \rho \times c^{3} t^{3} /\hbar
\end{equation}
For the total number of operations performed during the existence of the Universe $ N \equiv \nu t\approx \rho \times c^{3} t^{4} /\hbar $, we will reproduce the result obtained  above. This expression, obtained in the framework of Friedmann's cosmology, corresponds exactly to statement (2) of the holographic model of space-time foam. In fact, the density of the Universe is close to critical, and. Consequently,
\begin{equation} \label{ek1_37}
  N\approx \rho \times c^{3} t^{4} /\hbar \approx \frac{t^{2} c^{5} }{G\hbar } \approx \left(\frac{t}{t_{Pl} } \right)^{2} \approx \left(\frac{l}{l_{Pl} } \right)^{2}
\end{equation}
We now turn to the memory estimate of the Universe $I$, the maximum mount of information, which she can register. Since $I=S/k_{B} \ln 2$, the problem reduces to calculating the maximum entropy of the Universe. The maximum entropy will be achieved by converting all matter into radiation. Using well-known expressions for the entropy of black radiation \cite{87as}, we find \cite{87bs}
\begin{equation} \label{ek1_38}
   I\approx \left(\rho c^{3} t^{4} /\hbar \right)^{3/4} =\left(N\right)^{3/4} \approx \left(\frac{l}{l_{Pl} } \right)^{2} \approx 10^{90} \, \, bits
\end{equation}
To realize such a gigantic amount of memory, it is required to use all degrees freedom of matter filling the Universe.

Let us briefly dwell on the original idea of $S$. Weinberg \cite{weys} to take into account already at the level fundamental constants influence the dynamics of the Universe. For this he constructed the mass of a hypothetical particle, using in addition to the fundamental constants $\hbar$, $c$, $G$ also the Hubble parameter $H$,
\begin{equation} \label{ek1_39}
   m=\left(\hbar ^{2} H/Gc\right)^{1/3}
\end{equation}
As it turned out, this mass does not differ much from the mass of the “typical” elementary particle $m\approx 100MeV$. Note that in the presence of four initial constants $\hbar$, $c$, $G$, $H$ the mass construction procedure is not unambiguous.

Proper gravitational energy of such a particle
\begin{equation} \label{ek1_40}
  E_{g} =Gm^{2} /\left(\hbar /mc\right)=Gm^{3} c/\hbar
\end{equation}
Using expression (\ref{ek1_39}) as mass, we find
\begin{equation} \label{ek1_41}
   E_{g} =H\hbar
\end{equation}
Since $H\sim 1/T$, where $T$ is the life time of the Universe, the value of $E_g$ can be interpreted as minimum quantum of gravitational energy,
\[{E_{g}^{0} =H_{0} \hbar \approx 10^{-52} J,\quad H_{0}^{-1} \simeq 4.55\times 10^{17} \sec ,}\]
\begin{equation} \label{ek1_42}
   {M_{g}^{0} =\frac{E_{g}^{0} }{c^{2} } \simeq 10^{-66} \, \, g}
\end{equation}
The index "0" indicates the current values of the quantities. Given that according to SCM  in the energy budget of the Universe is currently dark energy dominates in the form of the cosmological constant $\Lambda$ ($\Lambda \approx 10^{-56} cm^{-2} ,\; \rho _{\Lambda } \approx 6\times 10^{30} g/cm^{3}$) we construct from the quantities $\hbar , c, G$ , and $\Lambda$ the quantities of the dimension of mass. In the result we obtain two significantly different masses
\[{m_{1} =\frac{\hbar }{c} \sqrt{\Lambda } \simeq 2\times 10^{-65} g,}\]
\begin{equation} \label{ek1_43}
 {m_{2} =\frac{c^{2} }{G} \frac{1}{\sqrt{\Lambda } } \simeq 10^{56} \, \, g}
\end{equation}
For a Universe with a dominant cosmological constant $H\sim c\sqrt{\Lambda}$. Therefore, $m_{1} \sim H \hbar$ and we can interpret the mass $m_{1}$ as the mass of quantum of gravitational energy (graviton).The mass $m_{2} \sim \Lambda H^{-3} \sim 1/\sqrt{\Lambda}$ can be interpreted as the mass of the observable part of the Universe (mass inside the Hubble sphere). In fact, represent this mass in form
\begin{equation} \label{ek1_44)}
  M_{g} =\frac{c^{2} }{G} \frac{1}{\sqrt{\Lambda } } =\frac{\Lambda c^{4} }{G} \frac{1}{\sqrt{\Lambda ^{3} } } \frac{1}{c^{2} }
\end{equation}
The first factor represents the energy density $\rho _{\Lambda}$ generated by the cosmological constant $\Lambda$, while the second factor represents the volume of the Hubble sphere. Representing $M_{g} =N_{g} m_{g}$, where $m_{g} = m_{1} = \frac{\hbar}{c} \sqrt{\Lambda}$ for the total number of gravitons in the observable Universe $N_g$, we find
\begin{equation} \label{ek1_45)}
   N_{g} =\frac{c^{3} }{G\hbar \Lambda }
\end{equation}
Interestingly, this number coincides, up to a numerical factor, with the ratio  of the Planck density $\rho _{Pl}$ o the density of the cosmological constant $\rho _{\Lambda}$ \cite{weys}
\begin{equation} \label{ek1_46}
   \frac{\rho _{Pl} }{\rho _{\Lambda } } \approx \frac{c^{7} }{\hbar G^{2} } \frac{G}{\Lambda c^{4} } =\frac{c^{3} }{\hbar G\Lambda } =N_{g} \approx 10^{120}
\end{equation}
In conclusion of this section, we will dwell on an important question - the possibility of experimental (observational) testing of the hypothesis of the existence of space-time foam. Planck length $l_{Pl} \sim 10^{-35} m$ is so small that we need an astronomical (even cosmological) distance $l$ in order for fluctuations $\delta l$  to become available to measurement. The hypothesis can be refuted in the case of observation of the so-called Airy pattern in images of distant sources obtained using space telescopes. Diffraction pattern that occurs when light passes through evenly illuminated round hole, has a bright area in the center, known as "Airy's disk". Overall diffraction pattern including spot and concentric bright rings represents  Airy pattern.

Consider light (with wavelength $\lambda$) from distant quasars or bright active galactic nuclei \cite{12s,13s}. Due to the quantum fluctuations of space-time, the wave front (flat in the classical case) becomes “foamy” due to random phase fluctuations $\Delta \varphi \sim 2\pi \delta l/l$ \cite{13s}.

The phase distortion is cumulative. Upon reaching the value $\Delta \varphi \sim \pi $, the wave front is distorted so much that its observation with the help of traditional interferometers becomes impossible.

Let us consider as an example \cite{13CD} the possibility of observing the active galactic nucleus PKS1413 + 135 \cite{pa} with redshift $z \sim 0.25$. The distance to this object $l \approx 1.2 \, G \, p \,c$ and observation is carried out for $\lambda =1.6 \,\mu m$.   Using this data, we find for $\delta l=l^{1-\alpha } l_{Pl}^{\alpha}$
\begin{equation} \label{ek1_47}
   \Delta \varphi \sim 2\pi \delta l/l\sim \left\{\begin{array}{l} {10\times 2\pi ,\quad \alpha =1/2\quad } \\ {10^{-9} \times 2\pi ,\quad \alpha =2/3\quad } \end{array}\right.
\end{equation}
Thus, the observation \cite{14s} by the Hubble Space Telescope of the Airy ring for active galactic nucleus PKS1413 + 135 \cite{pa} excludes the random walk model, but does not exclude the holographic model of space-time foam.

\section{Conclusion}

Historical experience teaches that the most interesting discoveries in physics occur when transition to new characteristic scales of quantities describing the problem under study. The boundary of the region in which a certain paradigm operates is determined by the fundamental  constants. For example, the transition from classical mechanics to relativistic is controlled by the speed of light $c$, and the transition from classical mechanics to quantum Planck's constant $\hbar$.

The transition to the Planck scale is much more complicated, both quantitatively (this area is separated from the current parameters by tens of orders of magnitude) and qualitatively. (is the concept of continuous space compatible  with quantum mechanics?). These difficulties make us look for work arounds ways to solve the problem.

In particular, as a preliminary stage, we can consider the phenomenology of. Phenomenology as a way taking into account the quantum manifestations of classical gravity allowed us to obtain two fundamental results: the minimum length and space-time foam. Within studying the phenomenology of  Planck scale physics we got the following new results:
\begin{enumerate}
  \item Using the maximum power for the transition from the set of fundamental constants $\left( {\hbar ,c,G} \right)$ to the set $\left( {\hbar ,c,\eta } \right)$ a system of modified Planck units was constructed. With such a transition, the numerical values of Planck units are preserved, however, new interesting possibilities for interpretation of  the results was obtained.
  \item Restrictions on the parameters of arbitrary quantum clocks are obtained.
  \item The  equivalence of the holographic principle and the principle of maximum force is proved.
\end{enumerate}

In conclusion, we briefly dwell on the immediate prospects for research in field of phenomenology of Planck scale physics. These studies can be divided pour into two groups that pursue different goals. The first group puts at the forefront experimental search for quantum manifestations of classical gravity, while how the second focuses on a more fundamental question: is there in general, such an object of study as quantum gravity. First, we discuss the planned experiments to search for quantum manifestations of classical gravity. These experiments until recently were considered impossible and belonged to the class of gedanken experiments. Astrophysical experiments are closest to implementation. Astrophysics presents a number of opportunities to make weak gravitational effects observable (at least in the near future). In particular, these experiments include:
\begin{enumerate}
  \item Use commutation effects. In this case, weak effects will accumulate over a long exposure time.
  \item Observation of ultrahigh-energy particles for which new effects may be measurable in the near future.
\end{enumerate}

The Universe provides us with both of these possibilities.

Special interest represent the planned experiments, that will shed light on the status of quantum gravity. Recently published two projects of such experiments \cite{zak1,zak2}, close in formulation, but differing in implementation. It is assumed that experiments will make it possible to determine whether two objects (two micro diamonds), between which there is only gravitational interaction, can be in an entangled state, i.e. in a state of quantum superposition.  If a entangled state will be realized, then, according to the authors, this will mean that the only force acting between them (gravity), represents quantum interaction.

However the problem of determining the status of quantum gravity is so complex that even a positive  experimental result will not lead to a final solution to the problem. F. Dyson, the most authoritative opponent of quantum gravity, so commented shaft planned experiments: "The proposed experiments certainly represent great interest, but it is not clear to me whether they will be able to resolve the issue of existence of quantum gravity. The question I asked is whether a separate graviton is observed, this is a different question and it may have a different answer".

\end{document}